\documentclass[]{JHEP}
\usepackage{epsfig}

\def\beq{\begin{equation}}
\def\eeq{\end{equation}}
\def\bea{\begin{eqnarray}}
\def\eea{\end{eqnarray}}
\def\bq{\begin{quote}}
\def\eq{\end{quote}}

\def\bq{\begin{quote}}
\def\eq{\end{quote}}

\def\ord{\cal O} 

\newcommand{\leqsim}{\,\raisebox{-0.6ex}{$\buildrel < \over \sim$}\,}
\newcommand{\geqsim}{\,\raisebox{-0.6ex}{$\buildrel > \over \sim$}\,}

\newcommand{\be}{\begin{equation}}
\newcommand{\ee}{\end{equation}}
\newcommand{\ba}{\begin{eqnarray}}
\newcommand{\ea}{\end{eqnarray}}
\newcommand{\nn}{\nonumber}

\newcommand{\ie}{\mbox{{\em i.e.~}}}
\newcommand{\eg}{\mbox{{\em e.g.~}}}

\newcommand{\eqr}[1]{eq.(\ref{#1})}

\def\half{\frac{1}{2}}

\def\lim{\mbox{{\bf L}} }

\def\Vp{V_{\parallel}}
\def\Vt{V_{\perp}} 
\def\vp{V_{\parallel}}
\def\vt{V_{\perp}} 
\def\wp{W_{\parallel}}
\def\wt{W_{\perp}} 
\def\Wp{W_{\parallel}}
 
\def\Rt{R_{\perp}}
\def\ep{\varepsilon}

\preprint{OUTP-00-09-P}

\title{Hagedorn inflation of D-branes}

\author{Steven A. Abel\footnote{Steven.Abel@cern.ch} 
\\LPT, Universit\'e Paris-Sud, Orsay, 91405, France.}
\author{Katherine Freese\footnote{ktfreese@umich.edu}\\
Randall Physics Laboratory, University of Michigan, 
Ann Arbor MI 48109-1120.
\\Max Planck Institut fuer Physik, Foehringer Ring 6, Munich, 
Germany.}
\author{Ian I.~Kogan\footnote{i.kogan@physics.ox.ac.uk}
\\Theoretical Physics, 1 Keble Rd, Oxford OX1 3NP, UK.}

\abstract{We examine the cosmological effects of the Hagedorn phase
in models where the observable universe is pictured as 
a D-brane. It is shown that, even in the absence of a
cosmological constant, winding modes cause a negative
`pressure' that can drive brane inflation of various types 
including both power law and exponential. 
We also find regimes in which the 
cosmology is stable but oscillating (a bouncing universe)
with the Hagedorn phase softening the singular behavior
associated with the collapse.}

\begin{document}

\section{Introduction}
\label{sec:intro}

Inflation \cite{guth} is a beautiful solution to
several difficult
problems in cosmology; the horizon problem, the flatness problem,
and the monopole problem. In field theory however, the standard way to obtain 
inflation is to add a positive cosmological constant (which has 
a negative pressure $p=-\rho)$. This ingredient is, without 
doubt, the least attractive feature of standard inflation 
and it is generally extremely difficult to control its adverse effects 
(\eg the graceful exit and moduli problems).
It is worth asking therefore
if there are other forms of matter that can have a negative
pressure and hence give an 
accelerating scale factor. In this paper we introduce 
a candidate that actually has this property -- open strings on D-branes
at temperatures close to the string scale. 

At sufficiently high temperatures and densities 
fundamental strings enter a curious `long string' 
Hagedorn phase~\cite{carlitz,general,deo,thorl,abkr,after}.
To date applications of this phase have been 
quite limited in string cosmology~\cite{general} because
the thermodynamics is governed by the finite temperature partition
function. A rigorous analysis
therefore requires nothing less than solving the string system 
in a cosmological setting, a difficult problem that might
at best be tractable only in a few special cases. 
Moreover, in order to understand the effect of macroscopic 
phenomena such as winding modes we need the microcanonical 
ensemble (as we shall discuss) -- 
an ensemble that does not particularly lend itself 
to cosmological applications. 

In this paper we show that, for certain systems, it is possible to 
bypass these technical difficulties by using a 
classical random walk picture to model the behaviour of the strings in 
cosmological backgrounds. The particular systems we will focus on 
are D-branes in the weak coupling limit~\cite{polch}. In this limit one can 
separate the energy momentum tensor into two components; a 
localized component corresponding to the D-brane tension, and 
a diffuse component that spreads into the bulk corresponding to 
open string excitations of the brane. (At the risk of causing confusion 
we will often refer to the latter as a `bulk' component.) We will 
in addition allow a bulk cosmological constant, although our focus in this 
paper will be on the cosmology when the combined effect of 
the brane and bulk cosmological constants is subdominant.

The random walk picture allows us a first glimpse
at the cosmological effects of a primordial Hagedorn phase of 
open strings on branes and we find
two interesting types of behaviour. 

\begin{itemize}

\item The first we call {\em Hagedorn inflation}. 
We will show that the transverse 
`bulk' components of a D-brane's energy-momentum tensor can be negative. 
If all of the transverse dimensions have winding modes,
this negative `pressure' causes the brane to 
power law inflate along its length with 
a scale factor that varies as $a\sim t^{4/3}$
even in the absence of a nett cosmological constant 
(as shown in eq.(\ref{61})).
If there are transverse dimensions that are large (in the sense that 
the string modes are not space-filling in these directions),
then we can find exponential inflation (as shown in eq.(\ref{expinf})). 

\item If there is a small but negative cosmological constant,
the Universe can enter a stable but oscillating phase; \ie
a `bouncing' universe. The nett effect of the Hagedorn phase 
is to soften the singular behaviour associated with the collapse. 
Such singularity smoothing is a familiar aspect of strings,  but 
the nice feature here is that we find it in a purely perturbative
regime. 

\end{itemize}

We should at this point also emphasize a 
general observation that we make, 
namely that the diffuse stringy component 
can have a dominant effect on the cosmology even 
in the weak coupling limit. At first sight this 
may be somewhat surprising given that the
intrinsic tension energy of a D-brane goes like
$\rho_{br}\sim 1/g_s \sim 1/\hat{\kappa} $ where $g_s$ is the string
coupling and $\hat{\kappa}$ is the effective gravitational coupling. 
However, we will see that the {\em cosmological} 
effects of the two components are proportional to 
$\hat{\kappa}^4 \rho_{br}^2 $ and $\hat{\kappa}^2 \rho$ for the 
brane and diffuse `bulk' components respectively.  Then
the contribution of the brane component is proportional to
$\hat{\kappa}^4 \rho_{br}^2 \sim \hat{\kappa}^2$.
Since  $\frac{1}{g_s}\geqsim 
\rho \geqsim 1$ to be in the Hagedorn phase, if \eg the transverse 
volumes are of order unity in string units, then the contribution of
the bulk component is
$\hat{\kappa} \geq \hat{\kappa}^2 \rho \geq \hat{\kappa}^2$. Hence
the cosmological effect of the diffuse bulk component  
can be dominant when $g_s$ becomes small.

We begin in sections 2 and 3 by deriving the energy-momentum tensor,
the principal ingredient of Einstein's equations.
Since the results can be understood rather intuitively, this 
part of the discussion is organized so that cosmologists 
(and indeed anybody else) can skip the bulk of 
sections 2 and 3 concerning string thermodynamics and proceed directly 
to the energy-momentum tensor which is summarized at the 
end of section 3.

The thermodynamic discussion of section 2 gives a detailed 
introduction to the behaviour of both 
type I and type IIA/B open strings on D-branes as calculated from the 
microcanonical ensemble in a flat background. 
In particular we discuss the importance of macroscopic 
modes such as winding modes. 
Much of this section is a collation of results from 
ref.\cite{abkr}. We then reintroduce the classical random walk picture paying 
special attention to the meaning of quantities such as average
string length.

In section 3
we use the thermodynamic results to calculate the
energy momentum tensor $T_{\mu\nu}$ of the Hagedorn phase.
$T_{\mu\nu}$ enters into the higher
dimensional Einstein's equations and determines the cosmology,
and in particular we show that open string winding modes 
gives negative transverse components. 
For convenience the results for $T_{\mu\nu}$ are summarized 
at the end of section 3 where we also discuss the heuristic interpretation
of this negative pressure.

Armed with the energy-momentum tensor,
we examine the resultant cosmology.
In sections 4 and 5, we solve the equations 
of motion with various ans\"atze, and find the advertised
inflationary behaviour as well as bouncing solutions  
with singularity smoothing behaviour.

We will, purely for definiteness, consider 
adiabatic systems in solving the evolution equations
of the universe.  Under the assumption
of adiabaticity, the inflationary growth period 
drives down the temperature of the system; eventually
the temperature drop causes
the universe to leave the Hagedorn regime, and consequently
inflation ends automatically. However adiabatic systems 
are probably unable to provide a realistic scenario 
with sufficient inflation.
In section 6 we therefore discuss how, in non-adiabatic systems,
inflation can be sustained. We conclude in section 7.

\section{The Hagedorn phase and random walks} 
 
This section presents some background 
thermodynamics needed to get $T_{\mu\nu}$
 (which enters into Einstein's equations) in the Hagedorn regime. 
In section 2.1 we review the thermodynamic properties of 
D-branes in toroidal compactifications. These compactifications 
allow us to use the {\em microcanonical} ensemble, which is 
defined in terms of global parameters such as total energy 
and volume.  The importance of the
microcanonical ensemble is that it allows a 
rigorous understanding of the effect 
of winding modes. This understanding enables us
in sections 2.2 and 2.3 to extend the analysis
to more general universes of any shape; in particular, we can
determine those cases in which thermodynamics and hence cosmology
can be studied.  Essentially, we will 
argue that thermodynamics only makes sense in a cosmological setting 
for those systems in which the canonical 
and microcanonical ensembles agree. Where this is not the case 
the systems are dominated by large scale fluctuations.

For the systems in which the two ensembles agree,
we will derive a expression for the 
partition function based on a heuristic random walk argument.
In particular this expression gives a geometric understanding of the 
partition function that allows us to discuss its validity in
limits of high energy density, small volume, {\em etc}. 

In section 3, we then use this partition function
to find the energy momentum tensor, $T_{\mu\nu}$.
Readers whose principal interest is cosmology
may wish to read subsection 2.1 and then 
proceed directly to the summary of the results for $T_{\mu\nu}$
in section 3.4.

\subsection{String thermodynamics and the Hagedorn phase}

The Hagedorn phase arises in theories containing fundamental
strings because they have a large number of internal degrees of 
freedom. Indeed, because of the existence of many 
oscillator modes, the density of states 
grows exponentially with energy $\varepsilon$, $\omega(\varepsilon) 
\sim \varepsilon^{-b}
e^{\beta_H \varepsilon}$, where the inverse Hagedorn temperature $\beta_H$ 
(where $\beta =1/T$)
and the exponent $b$ depend on the particular
theory in question (for example heterotic or type II)~\cite{carlitz}. 
For type I,IIA,IIB strings the numerical value of the 
inverse Hagedorn temperature
is $\beta_H=2 \sqrt{2} \pi$ in string units.
It is easy to see that thermodynamic quantities, such as the 
entropy, are liable to diverge at the 
Hagedorn temperature;
obtaining the partition function $Z$ with the
canonical ensemble and multiplying
by the usual Boltzmann factor $e^{-\beta \varepsilon}$,
one finds an integral for the partition function (for large $\varepsilon$)
$Z \propto \int d\varepsilon \varepsilon^{-b} e^{(\beta_H - \beta) 
\varepsilon}$ which diverges at
$\beta = \beta_H$ for $b \leq 2$. 

If an infinite amount of energy is required to reach $T_H$, then 
we say that the system is {\em limiting}. If not then 
the system is said to be {\em non-limiting}. 
Already the simple canonical ensemble above 
indicates that the Hagedorn temperature
is a limiting temperature for the $b\leq 2$ cases~\cite{general}.
The remaining systems seem to be non-limiting, and until 
recently it was thought that this might imply some kind
of Hagedorn phase transition (drawing strongly on the analogy with 
the quark-hadron phase transition) to more fundamental degrees of freedom.
However, in ref.\cite{abkr} it was shown that all string systems 
are in fact limiting, including 
arbitrary D$p$-branes (\ie open strings attached to $p$  dimensional 
defects). 

The limiting/non-limiting question is a rather
subtle one, and in order to resolve it one must
first allow for large scale fluctuations by using the 
microcanonical ensemble 
(as pointed out in the early papers of Carlitz and 
Frautschi~\cite{carlitz}) and second, retain
full volume dependence 
(as pointed out for heterotic strings by Deo et al \cite{deo} and 
for open strings in ref.\cite{abkr}).
Once both of these factors are included, it becomes clear that
in all cases the Hagedorn temperature
is truly a limiting temperature rather than an indication
of some sort of phase transition. 
Perhaps the best evidence for this 
is that the Hagedorn phase completes (by entropy matching) 
a phase diagram which includes other 
non-perturbative phases such as black holes. (There is a 
sense in which the entire Hagedorn phase can be thought of as 
a first-order phase transition from Yang-Mills/supergravity 
degrees of freedom
to black-branes/black-holes. We return to this point later.) 

Let us summarize the rigorous results from ref.\cite{abkr} for 
open and closed strings in a toroidally compactified space.  
The most direct route to the thermodynamics is to 
evaluate the one loop partition function with the Euclidean 
time coordinate, $\tau$, compactified with radius $\beta$.
However for the random walk discussion later, it is useful to 
begin with the density of states 
$\omega (\varepsilon )$ of a single string of energy $\varepsilon$
from which the same results are obtained;
\be
\label{omega} 
\omega({\varepsilon}) =
\left\{ 
\begin{array}{ll}
 \beta_H\, \frac{\Vp}{\Vt} \; V_o
 \frac{e^{\beta_H \,\varepsilon}}
{(\beta_H\,\varepsilon)^{\gamma_o+1}} & \mbox{ open}\nonumber \\
 \beta_H\, V_c
 \; \frac{e^{\beta_H\,\varepsilon}}{(\beta_H\,\varepsilon)^{\gamma_c+1}}
& \mbox{ closed.}
\end{array}
\right. 
\ee
In the above $\Vt$ and $\Vp$ are the volumes transverse and perpendicular 
to the D-brane. For open strings,
\be 
\label{gamma2}
\gamma_o =  {d_o \over 2} -1,
\ee
where $d_o$ is the number of dimensions transverse to the brane in 
which there are {\em no windings} and $V_o$ is the volume of this space
(if there are windings in all dimensions, then $V_o = {\cal O}(1)$ 
in string units).
Similarly, for closed strings,
\be
\gamma_c = \frac{d_c}{2}
\ee
where $d_c$ is the number of dimensions in which closed strings
have {\em no windings} and, again, $V_c$ is the volume of this space.
Note that $\gamma_o$ and $\gamma_c$ are $\varepsilon$-{\em dependent} 
critical exponents, because winding modes are quenched or
activated depending on the string energy. Below we shall use
$\gamma$ to stand for either $\gamma_c$ or $\gamma_o$ as
appropriate.

We now collect the results obtained in ref.\cite{abkr} in 
the {\em microcanonical ensemble}
working in an approximation to the thermodynamic limit. 
The two main types of behaviour are the single-long-string or 
non-limiting ${\bf NL}$ behaviour (see discussion below where we discuss
the consequences of placing `non-limiting' systems in thermal contact with
limiting systems), with entropy density
\be
\sigma \equiv S/\Vp
\ee 
of the form
\be
\label{nlen}
\sigma_{{\bf NL}[\gamma]} \approx \beta_H\,\rho-{1+\gamma \over \Vp}\,{\rm
log}\,(\rho)
,\ee
and the various types of `limiting' behaviour ${\bf L}[\gamma]$
\be
\label{len}
\sigma_{{\bf L}[\gamma]} \approx   
\left\{
\begin{array}{ll}
\beta_H\,\rho + 2\sqrt{\rho/ \Vt} 
\;\;\;\;\;\;&{\rm if}\;\;\;\;{\gamma=-1}
  \nn\\
\;\nn\\
\beta_H\,\rho \,+\,{\gamma-1\over \gamma}\,\rho^{\gamma\over \gamma-1}
\;\;\;\;\;\;&{\rm if}\;\;\;\;{\gamma =-\half,\half}
\nn\\
\,\nn \\
\beta_H\,\rho +  {\rm log}\,(\rho)   
\;\;\;\;\;\;&{\rm if}\;\;\;\;{\gamma=0}
\nn\\ 
\,\nn\\
\beta_H\,\rho -e^{-\rho}  
\;\;\;\;\;\;&{\rm if}\;\;\;\;{\gamma=1,}
\end{array}
\right.
\ee
up to positive constants of $\ord$ (1) in string units. 
Here 
\be
\label{eq:rho}
\rho \equiv E/\Vp \, ,
\ee
where $E$ is the total energy of the system. Note that the 
microcanonical ensemble results are expressed in terms of 
global parameters such as 
total energy which are valid in a toroidal compactification. 
We shall find that it is only possible to generalize the 
discussion for those systems in which the 
canonical and microcanonical results agree.

The {\bf L}[-1] system is our `standard' high energy regime. 
From eq.(2), we see 
that it corresponds to $d_o=0$, so that all dimensions have windings
and it is the system which is always reached provided that the volumes are 
finite and the energy density is high enough.
These systems are equivalent to $D-1$ branes
where $D$ is the total number of space-time dimensions
(in other words freely moving open strings). (Once there are many 
windings, we can 
T dualize the Dirichlet directions so that they become Neumann directions
much smaller than the string scale -- the winding modes become 
a spectrum of Kaluza-Klein modes indicating
open strings which are energetic enough to probe all of the $D-1$ Neumann
dimensions -- even the small ones.)

There are two other sorts of behaviour,
marginal limiting ${\bf ML}$ and 
weak limiting ${\bf WL}$ with entropy densities
\ba
\label{clen}
\sigma_{\bf ML} &\approx& \beta_H\,\rho -
\rho^{2D-3}\,(V_{D-1})^{2D-4}\,e^{-\rho\,R^{D-3}} \nn\\
\sigma_{{\bf WL}} &\approx& \beta_c\,\rho + {\beta_c  \,f\over 2} \,\rho
-{\beta_c^2\, \Vp\,f^2 \over 24} \, \rho^2
\ea
respectively, where $f=\Vp/\Vt$.

Whether a system is limiting or not is a function of the 
global parameters of the system such as total energy, $E$, and 
volumes $\Vt$,~$\Vp$. For example, imagine increasing the 
energy of an {\bf L}[-1/2] system. These systems are characteristic 
of an intermediate energy phase of open string systems possessing 
one large transverse dimension without windings. As the energy is increased 
windings will eventually be excited in this direction. 
If the transverse dimension is
of order $R_\perp$, this happens at an energy threshold $E=R_\perp^2$,
and for higher energies the thermodynamic behaviour changes to that of 
an {\bf L}[-1] system, the universal high-energy 
regime\footnote{
Note that the decoupling of winding modes implied by 
these equations of state is universal and 
in particular independent of the zero-temperature vacuum 
contribution to the 
partition function which may or may not be finite as the 
transverse radius becomes infinite~\cite{inprep}. This is because in the 
finite temperature piece of the partition function the winding 
modes are always accompanied by a Boltzmann-like suppression factor. 
A related point is that 
as well as the above contributions there is a suppressed
contribution to the entropy density of 
order $\sigma_{YM} \sim T^{D-2\gamma_o-3}$ 
 corresponding to  a Yang-Mills gas in the ${D-2\gamma_o-3}$
dimensions 
with windings and KK modes. In the Euclidean approach 
this contribution comes from a UV cut-off in the Schwinger integral
and represents a field-theoretic contribution which is subdominant for
temperatures close to $1/\beta_H$ but again is universal.}.

The {\bf NL} behaviour is so called because the temperature 
is higher than $T_H$
as can easily be seen when we calculate the temperature
of the subsystems from 
\ba
S_i(E_i)&=&\log \Omega_i(E_i)\nn\\
T_i^{-1}&=& \beta_i = \partial S_i /\partial E_i  ,
\ea
where $\Omega_i$ is the microcanonical density of states
in the subsystem $i$. The non-limiting systems formally obey
\be
T_{{\bf L}} < T_H
<T_{\bf NL}.
\ee
However it is important to realise 
that this simply means that the ${\bf NL}$ regimes 
are {\it transients} for a finite ten-dimensional volume.
Equilibrium can never be achieved when they are in thermal contact 
with the surrounding (colder) limiting system of closed strings.

We can compare the microcanonical ensemble results reviewed
in the previous two pages to the 
canonical ensemble                    by taking the thermodynamic limit
(\ie letting $V_{||}$ become infinite whilst keeping the
density on the brane $\rho$ constant).
If we consider strings attached to a single brane embedded in 
$D=10$ space time dimensions, 
then there is universal agreement between the canonical 
and microcanonical ensembles 
when there are $0<d_{\perp} <4$ transverse dimensions. 
In this case heading towards the thermodynamic limit 
quenches winding modes -- the transverse dimensions can
be thought of as effectively infinite in this limit. 
However when $d_{\perp}> 4$ (D$p$-branes in ten dimensions  with $p<5$),
modes as large as the transverse dimensions 
can be quenched if $\Vp \ll \Rt^2$, giving
${\bf NL}$ behaviour, or activated into a ${\bf L}[-1]$ system for
$\Vp \geqsim \sqrt{\Vt}$ (when the saddle
point approximation is valid). 
Taking $\Vp \rightarrow \infty $ with $\rho $ 
fixed and $\Rt$ constant leads to the onset of {\bf L}$[-1]$ behaviour
in the thermodynamic limit.
${\bf WL}$ behaviour occurs in an intermediate 
region of parameter space where windings are being quenched in 
an {\bf L}[-1] system and the usual saddle point approximation
is invalid. 
Finally closed strings in the thermodynamic limit have a critical
dimension $D=3$, where $D$ is the number of spacetime
dimensions. For $D\leq 3$, we get standard canonical behaviour, 
${\bf L}[(D-1)/2]$. On the other hand, for
$D>3$ winding modes generate `marginally limiting' 
behaviour ${\bf ML}$.

Some examples of different situations are shown (somewhat
impressionistically) in 
figure 1 where the dimension dependence is evident.
The entropic preference of strings for branes with higher dimensionality 
has the obvious interpretation that, when the strings are volume filling, the 
larger dimension branes `cut across' more strings.  

\FIGURE[htp]{
\hspace*{0.5in}
\epsfxsize=5.0in
\epsffile{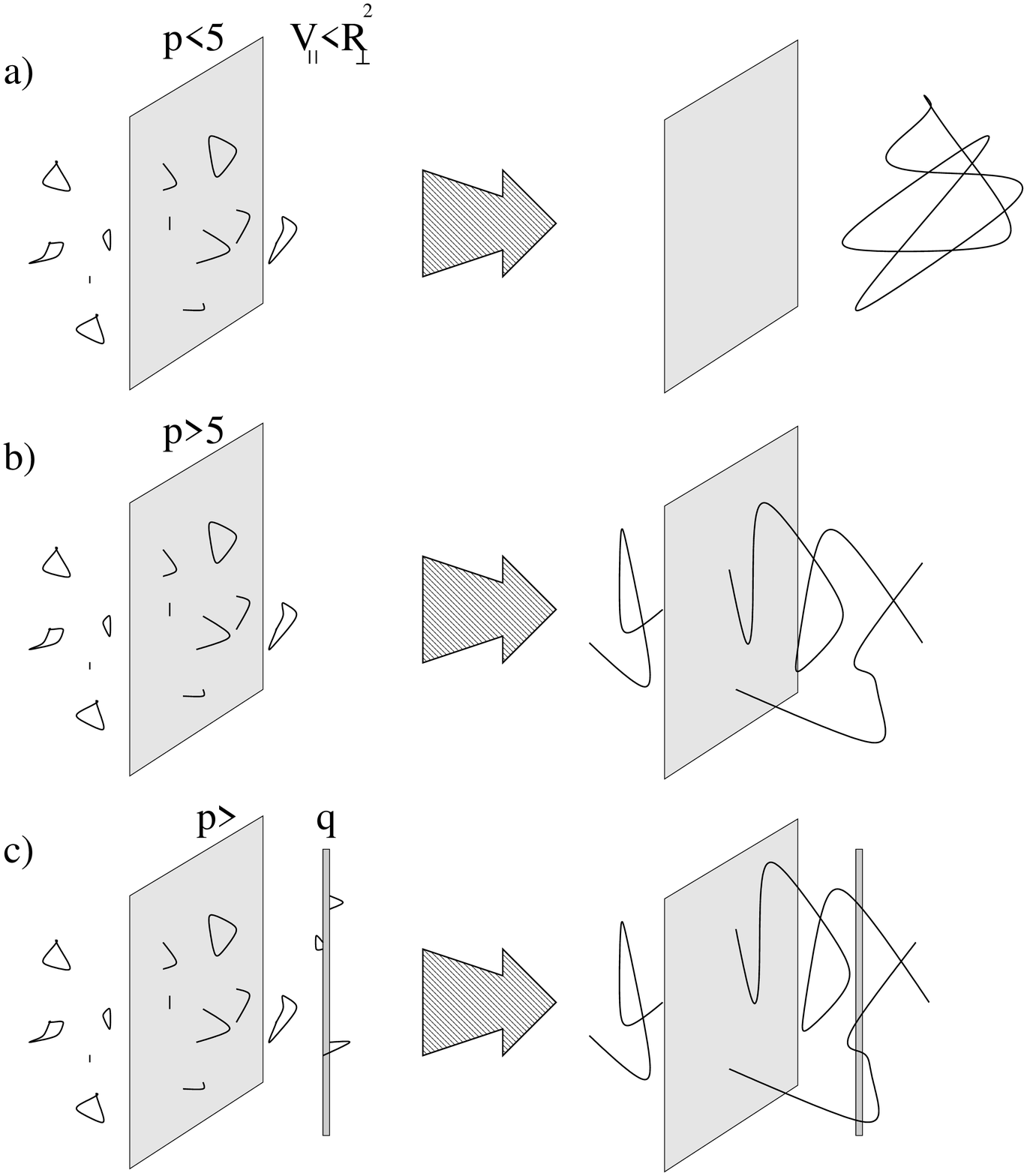}
\caption{ Examples of D-branes in thermal contact, 
showing (schematically) the direction of energy flow in order to achieve 
equilibrium. 
{\bf a:} A non-limiting (open string excitations of D-brane) 
plus marginal limiting (closed string) system. 
Energy is lost into the transverse volume until the energy density 
on the non-limiting system falls below the critical Hagedorn density. 
{\bf b:} A limiting (open string excitations of D-brane) 
plus marginal limiting (closed string) system. 
Energy flow is into open string 
winding modes as $T\rightarrow T_H$. 
{\bf c:} Limiting D$p$ brane plus limiting D$q$ brane plus marginal limiting 
(closed string) with $p>q$. 
Energy flows into open string windings on the 
D$p$ brane as $T\rightarrow T_H$.}}

Note that 
in the microcanonical ensemble we always 
use the definition of temperature
as derived from the microcanonical definition of entropy, $\beta = 
\partial_E S(E,V,N) $ where $E$ is the energy of the total system.
The discussion above illustrates 
the main reason for this rather convoluted 
set of definitions ($Z(\beta) \rightarrow \Omega 
\rightarrow S(E) \rightarrow \beta $);
the effect of winding modes 
cannot fully be explored using the canonical
ensemble. For example the canonical ensemble simply doesn't know about the 
{\bf ML} and {\bf WL} systems.
The microcanonical ensemble also allows us to look at
cases where parts of the system never come into equilibrium. 
This is true for instance for a D$3$-brane placed 
in a reservoir of closed strings. 
As we have seen, for large transverse volumes, 
a D$3$-brane formally has a temperature that is higher than $T_H$ 
and so any thermal energy that it may have quickly evaporates into the 
surrounding bath of closed strings. Only the
microcanonical ensemble allows one to compare the temperatures 
of {\em different} subsystems. Indeed the failure to do so leads to the false 
conclusion that the system {\em as a whole} might be non-limiting.
Moreover, strictly speaking 
the thermodynamic limit (\ie the assumption on which the 
canonical ensemble rests) does not actually exist for open strings. 
The phase diagram derived in ref.\cite{abkr} tells us that 
when trying to take the thermodynamic  
limit one inevitably encounters a black hole phase.
Thus it makes more sense to define the system with
global quantities such as total energy, and work in 
finite volumes taking an approximation to the naive 
thermodynamic limit if possible.

\subsection{The random walk interpretation}

Despite its many advantages the microcanonical ensemble
has one distinct disadvantage; 
in many cosmological backgrounds global properties such as 
total energy cannot be defined. 
We therefore need to be able to model these aspects of 
the thermodynamics with a more flexible  
approach. The random walk interpretation 
provides just such flexibility. For example, we can easily 
understand the {\bf NL}/{\bf L} cross over if 
we interpret a string of energy $\varepsilon $ as a 
random walk of length  $\sim \varepsilon$ in string units.
The size of such a random walk  
is $\sqrt{\varepsilon}$ so that, roughly speaking, to 
have modes of size $\Rt$ we require that  
the total energy $E=\rho \Vp \sim \Vp \sim \Rt^2 $.
In this classical picture the {\bf NL}/{\bf L} cross over occurs precisely 
when there are modes as large as the transverse volume\footnote{We
should caution against taking this picture too literally although,
as we are about to see, it does better than might be expected in 
explaining the thermodynamics. In particular the way to measure the 
`size' of the string is to put it in a box with a certain energy 
and see at what radius it begins to excite winding modes. This should
be borne in mind in order to avoid getting into circular arguments.}.
One helpful feature of the random walk interpretation 
is that we no longer have to bother 
about the precise topological properties of the 
space, merely its size in string units. On the other hand what we have 
gained from the rigorous analysis of the microcanonical ensemble 
is confidence in our understanding of which systems are stable,
when one can use naive {\em canonical} ensemble results 
(\ie as an approximation to the thermodynamic limit 
when the systems are limiting) and when the random walk interpretation
is valid.

An analysis of random walks was the approach taken by 
Lowe and Thorlacius for 
closed strings and Lee and Thorlacius for open strings attached 
to D-branes \cite{thorl}. These authors studied the Boltzmann
equations for the average number of interacting strings sections
of different lengths. The equilibrium solutions for
closed string loops of length $\ell$ for example show that
the (single string) density of states must be
of the form $ \omega(\varepsilon )\sim \frac{1}{\varepsilon} e^{\beta_H 
\varepsilon} $, where $\varepsilon = \sigma \ell $ and 
the temperature is related to the
 average total length $\bar L$ of string in the ensemble by
\be 
\label{eq:betah}
\beta_H= \beta - \frac{1}{\bar L}.
\ee

This supports the classical interpretation of the string
as a random walk with, $\varepsilon = \sigma \ell$, where 
$\sigma $ is the string tension encoding information about, for
example, the step-size.
In ref.\cite{abkr} it was shown that such a 
random walk interpretation also accounts for the volume dependence 
as follows.
First consider the distribution function $\omega (\ep)$ for
closed strings in  $D$ large 
space-time dimensions. The energy $\ep$ of  the
string is proportional to the length of the random walk. The number
of walks with a fixed starting point 
 and a given length $\ep$ grows exponentially as $\exp
\,(\beta_H\,\ep)$.
 Since the walk must be closed, this overcounts by a
factor of the volume of the walk, which we shall denote by 
\begin{equation}
 V({\rm walk}) = W \, .
\end{equation} 
Finally, there is a
factor of $V_{D-1}$ from the translational zero mode, and a factor of
$1/\ep$ because any point in the closed string can be a starting point.
The end result is
\be
\label{rw}
\omega(\ep)_{\rm closed} \sim V_{D-1}\cdot {1\over \ep} \cdot
{e^{\beta_H\,\ep} \over W}.
\ee
Now, the volume of the walk is proportional to $\ep^{(D-1)/2}$ if it is
well-contained in the volume $(R\gg \sqrt{\ep})$, or roughly $V_{D-1}$
if it is space-filling $(R\ll \sqrt{\ep})$. From here we get the
standard result \cite{general,deo}. We  have
\be
\label{eq:noncompact}
\omega (\ep)_{\rm closed}
 \sim V_{D-1} \,{e^{\beta_H \,\ep} \over \ep^{(D+1)/2}}
\ee
in $D$ effectively non-compact space-time dimensions, and 
\be
\omega (\ep)_{\rm closed} \sim {e^{\beta_H \,\ep} \over \ep} 
\ee
in an effectively compact space.     
Note that these results agree with those of eq.(1);
for example to obtain the same result as \eqr{eq:noncompact},
we take $d_c = D-1$, the number of space dimensions.

We can customize this analysis for open strings attached to 
a brane by a slight modification of the combinatorics.
(The more general case of intersecting $p,q$ branes was 
discussed in ref.\cite{abkr}). 
The leading
exponential degeneracy of a random walk of length $\ep$ with a fixed
starting point in say the D$p$-brane is the same as for closed
strings: $\exp (\beta_H\,\ep)$.
Fixing also the end-point at a {\it particular} point of the D$p$-brane
requires the factor $1/W$ to cancel the overcounting, just as in the
closed string case. Now, both end-points  move freely   
in the part of the brane occupied by the walk. This gives
a further degeneracy factor
\be
\wp^2
\ee
from the positions of the end-points.    
Finally, the 
overall translation of the walk in the excluded transverse volume gives a
factor
$\Vp / \wp$. The final result is: 
\be 
\omega (\ep)_{\rm open} \sim 
{\Vp \over \wt} \;   
\exp\,(\beta_H\,\ep) .
\ee 
Thus we see the sensitivity of the density of states to the 
effective volume of the random walk in the transverse directions. If the walk 
is well-contained in these directions, $(R_{\perp} \gg \sqrt{\ep})$, then  
we find $\wt \sim \ep^{d_{\perp}/2} $ and  
\be
\omega (\ep)_{\rm open} \sim {\Vp \over  
\ep^{d_\perp /2}} \;\exp\,(\beta_H\,\ep) \,\,\hspace{1cm} 
R_{\perp} \gg \sqrt{\ep} \, .
\ee 
On the other hand, if it is space-filling in the transverse 
directions $(R_\perp \ll
\sqrt{\ep})$, the transverse volume of the walk is just $\wt \sim 
\Vt$ and we find  
\be 
\omega(\ep)_{\rm open} \sim {\Vp \over \Vt}  
\;\exp\,(\beta_H \,\ep) \,\,\hspace{1cm} 
R_{\perp} \ll \sqrt{\ep} ,
\ee 
in agreement with eq.(\ref{omega}). 

\subsection{Random walks in a cosmological background}

The above thermodynamics involved a Euclidean metric. 
How can we adapt the results for a
cosmological background? First we assumed that the density 
of states increased as $\exp (\beta_H\,\ep)$. In a sense the parameter 
$\ep$ is now no longer the energy but becomes merely something to be 
integrated over. In the case of a non-trivial 
metric the most natural interpretation of $\ep$ is that it is 
the proper length of 
a string in the bulk and certainly we can always go 
to the local inertial frame in which a small portion of the string 
has the usual Euclidean energy~$\equiv$~length equivalence. 
In addition it is clear that the average number of 
strings of a given proper length is well defined and so
the arguments of Lee, Lowe and Thorlacius then give us 
the correct distribution of proper string lengths in equilibrium.
Furthermore eq.(\ref{eq:betah}) then gives us a working 
definition of temperature in terms of average proper string lengths.

To put these arguments on a firmer footing we now derive 
one aspect of the strings' behaviour in a cosmological 
background that will be useful in the discussions 
that follow; namely 
that in a slowly varying background a classical string
of proper length $\ell \sim \varepsilon$ occupies 
a volume of proper-size $\sqrt{\varepsilon}$.

Consider a classical string of proper length $\ell = 
\varepsilon/\sigma $ (where $\sigma \sim 1$ in string units) 
with one end point fixed in a $D-1$ dimensional space.
From our observations above we would expect such a 
string to have a density of 
states $\omega(\varepsilon)\sim e^{\beta_H\varepsilon}$. 
The crucial point is that we can  
arbitrarily divide this string into $N$ small strings 
each of which has one end free and one end that is fixed 
to the end of the previous string. Consequently 
the density of states of the large single string is the same as 
that of $N$ small strings each of which has one end fixed and 
with energies 
obeying  $\sum_i \varepsilon_i = \varepsilon$
where $\varepsilon_i$ is the energy of each string in its
local inertial frame. 
By choosing a sufficiently large $N$ we may always 
use the flat space approximation to evaluate the density of 
states of each small string rigorously (with one end fixed and
one end free), 
\be 
\omega_i (\varepsilon_i)=e^{\beta_H \varepsilon_i}.
\ee
In particular we find the same $\beta_H$ for all the 
strings. The total density of states is then given by 
\be 
\omega(\varepsilon) = 
\int \prod_i \left(
\omega_i d\varepsilon_i \right)
\delta (\varepsilon-\sum_i
\varepsilon_i ) = e^{\beta_H\varepsilon },
\ee
as required. 

It is now clear how to find the region occupied by the 
string. We measure this by determining the {\em gyration},
defined as the average size of the fluctuations 
of the free end from the fixed end, measured along null geodesics
passing through the latter. Thermodynamics in the local inertial frames
indicates that each small string has a spherically symmetric gyration 
with a radius $\sqrt{\varepsilon_i}$. To find the gyration of the 
large string we must (since they are average fluctuations) 
add those of the subsystems in quadrature.
Hence, regardless of which geodesic we choose,
the combined system has a gyration 
$\sqrt{\sum_i \varepsilon_i} = \sqrt{\varepsilon}$. 

Proceeding now to the volume dependence
of $\omega(\varepsilon)$,  
we first make the usual quasi-equilibrium 
approximation that equilibrium is established much more 
quickly than 
any change in the metric so that 
the metric may be taken to be approximately constant
in time when evaluating properties such as density.
In order to simplify matters, we also make the
(by now) familiar assumption that
the metric is factorizable into parallel dimensions $x$ and 
transverse ones $y$  
\be 
ds^2 = -n^2 dt^2 + g_{||ij}dx^idx^j +g_{\perp nm}dy^ndy^m ,
\ee
with the brane lying at $y^n=0$.
We define parallel and total volumes as
\ba 
\label{eq:parperp}
V_{||}(y') &=& \int dx dy \delta (y-y') \sqrt{g_{||}} \nonumber \\
W_{||}(y') &=& \int dx dy \delta (y-y') \sqrt{g_{||}} \eta (y,x) \nonumber \\
V &=& \int dx dy \sqrt{g_{||}}\sqrt{g_\perp } \nonumber \\
W &=& \int dx dy \sqrt{g_{||}} \sqrt{g_\perp } \eta (y,x)
\ea
where $\eta$ is a function that is one in the region 
of the random walk and zero elsewhere, and where 
$g_{||}$ and $g_\perp$ are the 
determinants of the metric in the brane dimensions and 
transverse dimensions respectively.
Note that we are using $x$ and $y$ as shorthand for
all the parallel and transverse coordinates. 
We also define
an averaging over the extra dimension with an overbar, 
\be 
\label{eq:avgquant}
\overline{O} = \frac{\int dy \sqrt{g_\perp} O(y) }{\int dy'
\sqrt{g_\perp } }\, .
\ee
For notational convenience we 
treat $\sqrt{g_\perp(y) }$ as a `$y$-dependent transverse volume',
\be
\label{eq:totalperp}
\Vt(y) = \sqrt{g_\perp(y) } \int dy' ,
\ee
so that the actual transverse volume is written 
\be 
\overline{\Vt}=  \int dy \sqrt{g_\perp(y) }.
\ee
Note that $V=\overline{\Vp}\overline{\Vt}$.
From the discussion above, we have $W_{||} = \varepsilon^{d_{||}/2}$
reflecting the fact that the string has typical size $\sqrt{\varepsilon}$. 
For example if we take a slice at $y=y'$ this should be 
\be 
W_{||} = \varepsilon^{d_{||}/2} = 
\int dx dy \delta (y-y') \sqrt{g_{||}} \eta =
\int dx \sqrt{g_{||}}(y') \eta (y')
\ee
so that $\eta $ must compensate for $\sqrt{g_{||}} $ in order to 
make this volume independent of $y'$. 

In order to examine the thermodynamics we now need to 
decide when we can use the 
microcanonical results. We first divide the parallel dimensions 
into small locally flat patches with $\sqrt{g_\perp } $
approximately independent of $x$.  In each patch it is consistent to use the 
microcanonical results, provided {\em there are no long range
fluctuations}. 
Thus we are generally prohibited from examining the {\bf NL} systems
since adjacent patches will not be in equilibrium.
\footnote{Also note that by examining the thermodynamics in a 
small region we do not artificially quench 
Kaluza-Klein modes as long as the local patches are much larger than
the string scale. This is not the case in the perpendicular 
directions however because of the presence of macroscopic winding
modes. Hence in a local patch with average volume $\overline{\Vp}$ 
we may define an energy density by for example 
$\rho=E/\overline{\Vp}$ where $E$ is the total energy in the patch.
However, for the thermodynamics, 
we must always include the whole transverse volume $\overline{\Vt}$ 
to avoid artifically {\em un}quenching winding modes.}

The random walk argument proceeds as before with  
one change. The factor that corrects for the translation of
the walk in the excluded volume is not $\Vp/\Wp$ but rather
$V/W$ for the {\bf L}[-1] system.
In a toroidal compactification these are of course the same
for the {\bf L}[-1] case. 
Here however, we must take account of the fact that the walk is 
squeezed by the metric so that it may have more `room' at one 
side of the compactified dimension than the other. 
This factor can also be written as 
\be 
\frac{V}{W} = \frac{\overline{\Vp}}{\Wp},
\ee
and we now find, for example, that 
the density of states in the {\bf L}[-1] system is 
\be 
\omega(\varepsilon ) 
\sim 
\frac{ \overline{\Vp }}{\overline{\Vt}} e^{\beta_H\varepsilon},
\ee
\ie the same expression as in the flat space case 
but with all volumes averaged over transverse dimensions as in 
eqs.(\ref{eq:avgquant}) and (\ref{eq:totalperp}). 

It is now possible to find $\log Z(\beta, \overline{\vp},\overline{\vt})
$ from $\omega (\varepsilon )$ by integrating over 
$\varepsilon $ with a Boltzmann weighting, which we 
do for the various sytems in the following section. 
Note that the $\beta $ appearing in the resulting partition 
function is the random walk definition of `inverse temperature' 
whose precise physical interpretation is given by eq.(\ref{eq:betah}). 

\section{Stress-energy tensor $T_{\mu\nu}$ in a bulk Hagedorn phase}

We now use these thermodynamic results to 
find the bulk energy momentum tensor during the Hagedorn regime.
We may find the energy momentum tensor from 
\be 
\label{eq:tuv}
\langle T^\mu_\nu \rangle = 2 \frac{g^{\mu\rho}}{\sqrt{g}} 
\frac{\delta \log Z(\beta,\overline{\Vp},\overline{\Vt}) } 
{\delta g^{\rho\nu}},
\ee  
where $\beta,~\overline{\vp},~\overline{\vt}$ are given by 
eqs.(\ref{eq:betah}),(\ref{eq:parperp}),(\ref{eq:avgquant}),(\ref{eq:totalperp}).
We will treat the functional derivative with respect to $g_{\mu\nu}$
in the following way.  We assume that small changes in the metric 
correspond to making small changes in the volumes in $\log Z$, 
\eg for a single extra dimension
\be
\label{eq:fnal}
{\delta Z \over \delta g_{55}} = \int dx' {\delta Z \over \delta
\overline{V}_\perp(x')} 
{\delta \overline{V}_\perp (x') \over \delta g_{55}} \, .
\ee
Then, in the case of only one extra dimension, we can write 
\be
\label{eq:fnal2}
\overline{\Vt} = \int dy \sqrt{g_{55}} =
{\int d^5x \sqrt{g_{55}} \over \int d^4x}
= {1 \over v\beta} \int d^5x \sqrt{g_{55}} \, ,
\ee
and 
\be 
\label{eq:fnal3}
{\delta \overline{\Vt} \over \delta g_{55}} = \frac{1}{  2
\sqrt{g_{55}}  v\beta} \, .
\ee
Then from eqs.(\ref{eq:fnal}--\ref{eq:fnal3}) we can determine
the functional derivative in eq.(\ref{eq:tuv}).
Our ansatz automatically means that $T_{05} = 0$
and hence $G_{05} = 0$; in other words we are not considering energy exchange
between the brane and the bulk. (In general there might
be energy flux between the two.)

We begin by evaluating the energy momentum
tensor for our `standard' case of ${\bf L}[-1]$
(windings in all transverse dimensions)
appropriate to high energies and densities;
the resulting $T_{\mu\nu}$ is presented in eq.(\ref{eq:negpres}). 
We then proceed
to ${\bf L}[\gamma \neq -1]$ cases (in which windings are quenched in
some of the transverse dimensions) and present the resulting
$T_{\mu\nu}$ in eq.(\ref{emgamma}).

\subsection{$T_{\mu\nu}$ at high energies and densities; {{\bf L}[-1]}}

We now apply eq.(\ref{eq:tuv}) to
the very high energy regime in which there are modes
stretching across the whole space in all transverse dimensions. 
In the nomenclature of ref.\cite{abkr} these are the limiting 
{\bf L}[-1] systems since $d_o=0$.

We first need the principal cosmological ingredient, 
the partition function.
The expression for the density of states in the {\bf L}[-1] system is 
\be 
\omega(\varepsilon ) 
\sim 
\frac{ \overline{\Vp }}{\overline{\Vt}} e^{\beta_H\varepsilon},
\ee
which, when integrated with a Boltzmann weighting, gives 
\be 
\label{eq:partition}
\log Z(\beta,\overline{\Vp},\overline{\Vt} ) = 
2 \frac{\overline{\Vp}^2 \beta_H^2}{\overline{\Vt} (\beta^2-\beta_H^2) } 
+ a_c \overline{\Vp} - \overline{\vp} 
\rho_c \frac{(\beta^2-\beta_H^2)}{2\beta_H} ,
\ee
where $a_c$ and $\rho_c$ are a critical pressure and 
energy density (defined with reference to the brane dimensions, 
\ie with dimensions $E_c/\overline{\Vp}$)
which are of order unity in string units~\cite{abkr}.
Here, subscript-$c$ refers to critical quantities to remain in the
Hagedorn phase.
These are the successive terms in a saddle point approximation.

After doing the functional derivative we 
can replace $\beta $ by using the saddle point result for the {\bf L}[-1] 
system,
\be
2 \frac{\overline{\Vp} \beta_H^2}{\overline{V_\perp} (\beta^2-\beta_H^2) }
\approx \sqrt{ 
\frac{\overline{\Vp}}{\overline{V_{\perp}} }
\beta_H (\rho -\rho_c)} ,
\ee
where as before, $\rho=E/\overline{\Vp}$.
For the saddle point approximation to be valid 
we require~\cite{abkr}
\be 
(\rho-\rho_c) \gg \Rt^4.
\ee
This is also the condition for 
the leading (first) term in $\log Z$ to dominate over the $a_c$ term, and 
for the $a_c$ term to dominate over the $\rho_c$ term. 
We neglect higher order terms.

In order to write the resulting energy-momentum tensor,
we define the parameters
\be 
\alpha_{||} = \frac{ \overline{\Vp} }{\Vp}\mbox{\hspace{0.4cm};\hspace{0.4cm}}
 \alpha_\perp = \frac{ \overline{\Vt} }{\Vt}
\ee
and a pressure 
\be
\hat{p}_{-1}= \frac{\sqrt{\rho }}{\overline{\Vt}^{3/2} }.
\ee
In the following section we shall argue that $ \alpha_{||} \approx 1$
wherever the Hagedorn phase is the dominant phase. 
If the off-diagonal components of the metric are small, then 
the variation with respect to $g^{\mu\nu}$ gives the 
bulk $\langle \hat{T}_{\mu\nu} \rangle $ terms; 
\ba
\label{eq:negpres}
\hat{T}^0_0 &=& - \hat{\rho} \, = \,
- \frac{\rho \alpha_{||}\alpha_\perp }{\overline{\Vt}}
\nonumber \\
\hat{T}^i_i &=& \hat{p}_{-1} \nonumber \\
\hat{T}^m_m &=& -\hat{p}_{-1}(2 \alpha_{||}-1) \, \approx \, -\hat{p}_{-1} ,
\ea
where $i$ labels the parallel $x$ directions and $m$ labels the 
transverse $y$ directions. 
We have dropped the $\langle \rangle$ notation for
the expectation value of the energy momentum tensor,
but shall continue to use hats to 
signify bulk properties such as the bulk density above, $\hat{\rho}$. 

It is instructive to consider the  
conservation equations derived 
from the Bianchi identity $T^{\mu\nu}_{;\mu}=0$. 
For example, restricting ourselves to 
the often discussed scenario with one extra 
dimension, $y$, and metric of the form 
\be 
\label{simple}
ds^2 = -n^2 dt^2 + a^2 dx^2 + b^2 dy^2 ,
\ee
the E-M conservation equations may then be written (we are using the
same notation as in ref.\cite{KKOP})
\ba
\frac{d \hat{T}^0_0}{dt} + (\hat{T}^0_0-\hat{T}_i^i) 3 \frac{\dot{a}}{a} 
+ (\hat{T}^0_0-\hat{T}_5^5)\frac{\dot{b}}{b} &=& 0 \nonumber \\
 \frac{d \hat{T}^5_5}{dy} + \hat{T}^5_5
\left(\frac{n'}{n}+\frac{a'}{a} \right)
+ \frac{n'}{n} \hat{T}_0^0 - 3 \frac{a'}{a} \hat{T}_i^i &=& 0.
\ea
We shall see shortly that $\alpha_{||} , \alpha_\perp \approx  1$  
so the first equation is a conservation law for entropy;
\be 
\frac{d(\sigma \overline{\Vp})}{dt}
\left( 1+
\sqrt{\overline{\Vt}/\rho}
\right) 
- 2\sqrt{\overline{\Vt}/\rho }.
\frac{d}{dt}\left(\overline{\Vp}\sqrt{\rho/\overline{\Vt}}\right)
\ee
where 
\be 
\sigma = \rho + 
2 \sqrt{\rho /\overline{\Vt}}.
\ee
Recall from ref.\cite{abkr} that the saddle point dominance condition is
\be
\sqrt{\overline{\Vt}/\rho} \ll \frac{1}{R_\perp^2}
\ll 1 .
\ee 
Hence to order $1/R_{\perp}^2$ we simply recover 
the expected entropy conservation law, $\frac{d(\sigma 
\overline{\Vp})}{dt}=0$.
Note that the entropy is evenly distributed because 
of the dominant first term; indeed to first order
$\hat{T}^0_0=\hat{\sigma}(y) $ where $\hat{\sigma}$ is the 
local entropy density in the bulk. 

\subsection{$T_{\mu\nu}$  for the {{\bf L}[$\gamma \neq -1 $]} systems.}

As the energy density of an {\bf L}[-1] system falls 
below certain energy thresholds, the modes that are 
sensitive to the size of the transverse dimensions (\eg winding 
modes in toroidal compactifictions) become quenched. Once this happens, 
the compactness of 
the quenched directions is of no further consequence for the 
thermodynamics, which aquires a different critical exponent, $\gamma$,
and can even temporarily become {\bf NL} (although the name `non-limiting' 
is probably misleading for the reasons discussed in the Introduction
and in ref.\cite{abkr}). If there are many transverse dimensions, 
and they are anisotropic, there may be a few energy thresholds and 
$\gamma$ may assume several (increasing) values before the density 
drops below the Hagedorn density $\rho_c$ and the system finally 
leaves the Hagedorn phase and enters 
the Yang-Mills phase.

In order to discuss the thermodynamics of these intermediate systems, 
we first introduce the function $\eta'(y)$ which gives support 
in the regions 
occupied by the random walking strings in the various transverse dimensions. 
(Note that $\eta(x,y)$ described the shape of a single string whereas 
$\eta'(y)$ is for all strings, so $\eta'(y)=0$ indicates 
that strings do not extend this far into the bulk.)

We define an averaging over the region given support by $\eta'$;
\be 
\label{eq:ep}
\overline{O}_{\eta'} = 
\frac{\int dy \eta' \sqrt{g_\perp} O(y) }{\int dy' \eta' \sqrt{g_\perp } }.
\ee
The single string density of states density of states now involves the 
ratio of the volumes 
\ba
V_{\eta'} &=& \int dx dy \eta' \sqrt{g_\perp} \sqrt{g_{||}} \nonumber \\
W &=& \int dx dy \eta'\eta \sqrt{g_\perp} \sqrt{g_{||}} 
\ea
(where $\eta$ is defined below eq. (\ref{eq:parperp})) so that 
\be 
\frac{V_{\eta'} }{W}= \frac{\overline{V_{||}}_{\eta'}}{W_{||}}.
\ee
The expression for the single string density of states 
becomes 
\be 
\omega(\varepsilon ) 
\sim \frac{\overline{V_{||}}_{\eta'}}{\overline{V_{\perp}}_{\eta'}} 
\frac{e^{\beta_H\varepsilon}}
{\varepsilon^{\gamma}}
\ee
where in accord with our previous 
definition of $\eta$, we have assumed that the strings occupy a 
proper volume $\varepsilon^{d_0/2}$ in the $d_0$ dimensions
where the modes do not fill the whole of space. 
As before we have $\gamma_0=d_0/2 -1$.

Defining 
\be
\overline{f}=\frac{\overline{V_{||}}_{\eta'}}{\overline{{V_\perp}}_{\eta'}} 
\ee
we may now use the expressions for the singular part of $\log Z$ 
found in ref.\cite{abkr};
\be 
\log Z_{sing} \sim
\left\{
\begin{array}{ll}
\Gamma(-\gamma) \overline{f} (\beta-\beta_c)^\gamma, & \gamma
\notin
\mbox{\bf Z}^+\bigcup \,\{0\}  \nn\\
&\nn\\
\frac{(-1)^{\gamma+1}}{\Gamma(\gamma+1)}
 \overline{f} (\beta-\beta_c)^\gamma \log (\beta-\beta_c ) & \gamma 
\in
\mbox{\bf Z}^+\bigcup \,\{0\}. 
\end{array}\right.
\ee
In the microcanonical ensemble, 
these expressions gives us the same relations as in eq.(\ref{len})
with $\rho $ replaced by 
\be 
\overline{\rho} = \frac{E}{\overline{\Vp}_{\eta'}},
\ee
whence, substituting the appropriate 
expression for $\overline{\rho} $ from eq.(\ref{len}) for each 
$\gamma$, we find the following for the 
{\bf L}[$\gamma\neq -1$] systems;
\ba
\hat{T}^0_0 &= & -\eta' \hat{\rho}
\nonumber \\
\hat{T}^i_i &=& \eta'\hat{p}_{\gamma} \nonumber \\
\hat{T}^m_m &=& \left\{
\begin{array}{ll}
-\eta'\hat{p}_{\gamma}\left( 2 
\alpha_{||}-1\right) \,\approx\, -\eta'\hat{p}_{\gamma} & \mbox{windings}
\nn\\
0 & \mbox{no windings,}
\end{array}\right. 
\ea
where
\be
\hat{p}_{\gamma} =   
\left\{
\begin{array}{ll}
\frac{1}{\overline{\Vt}_{\eta'}^{3/2}} 
\overline{\rho}^{\frac{\gamma}{\gamma-1}}
\;\;\;\;\;\;&{\rm if}\;\;\;\;{\gamma=-\half,\half}
  \nn\\
\;\nn\\
\frac{1}{\overline{\Vt}_{\eta'}^{3/2}} 
\log\overline{\rho}
\;\;\;\;\;\;&{\rm if}\;\;\;\;{\gamma=0}
\nn\\
\,\nn \\
\frac{1}{\overline{\Vt}_{\eta'}^{3/2}} 
e^{-\overline{\rho}}
\;\;\;\;\;\;&{\rm if}\;\;\;\;{\gamma=1.}
\end{array}
\right.
\ee
Because, as we see later, $\alpha_{||,\perp}\approx 1$, 
we shall henceforth drop the overbar notation.

\subsection{ 
{{\bf NL}[$\gamma $], {\bf WL} { and} {\bf ML} { systems.}}}

For the {\bf NL} systems, $\gamma > 1$ and
the microcanonical calculation tells us that 
\be
E=-\frac{(1+\gamma)}{\beta-\beta_c}. 
\ee
In these cases we have already argued that they cannot be 
in equilibrium with the remaining systems since 
the temperature that we formally derive from the 
microcanonical ensemble has $T>T_H$.
Indeed eq.(\ref{eq:betah}) also indicates 
the average total length of strings in the ensemble cannot be 
positive. The fact that there is no equilibrium solution to the 
Boltzmann equations is merely a different way of 
seeing that these systems are transient.
Little more information can be gained from the random walk 
picture in this case. However it seems likely that, 
as in the flat space microcanonical calculation, these 
systems tend to lose energy to closed strings in a
cosmological setting.

The {\bf WL} systems correspond to open strings
that do not satisfy the saddle point condition. This
is a cross over region where winding modes 
are just beginning to be excited, and the corresponding specific 
heat is therefore small (hence `weak limiting'). Unfortunately 
this case is also difficult to analyse because the microcanonical 
ensemble does not agree with the canonical result (indeed there {\em is} 
no canonical equivalent of these systems). Consequently it is difficult 
to find anything that we may interpret as 
$\langle T_\mu^\nu\rangle$ 
for these cases. The same is true for the {\bf ML}
systems which correspond to closed strings in the Hagedorn regime. 
Fortunately all of these systems have lower entropy than the 
limiting systems and can be neglected in the cosmology.

\subsection{Summary and discussion of $T_{\mu\nu}$}

We now summarize the results for the energy-momentum tensor.
We first drop the overline notation of the previous section 
and simply redefine $\Vt$ and $\Vp$ to be the transverse and parallel 
volumes covariantly averaged over the region
of the transverse dimensions covered by the strings.
(See eqs.(\ref{eq:parperp}),(\ref{eq:totalperp}),(\ref{eq:ep}).)
We define an energy density $\rho$ of strings 
\be 
\rho = \frac{E}{\Vp}.
\ee
For the limiting systems with critical exponent $\gamma$, {\bf L}[$\gamma$], 
we find that the `bulk' components of 
the energy momentum tensor are given by
\ba
\hat{T}^0_0 & = & -\hat{\rho}
\nonumber \\
\hat{T}^i_i & = & \hat{p}_{\gamma} \nonumber \\
\hat{T}^m_m &\approx & \left\{
\begin{array}{ll}
-\hat{p}_{\gamma} & \mbox{transverse with windings}
\\
0 & \mbox{transverse without windings,}
\end{array}\right.
\label{emgamma}
\ea
where $\hat{\rho}=\rho/\Vt$ and 
\be
\label{eq:phat}
\hat{p}_{\gamma} =   
\left\{
\begin{array}{ll}
\frac{1}{\Vt^{3/2}} 
\rho^{\frac{\gamma}{\gamma-1}}
\;\;\;\;\;\;&{\rm if}\;\;\;\;{\gamma=-1,-\half,\half}
  \nn\\
\;\nn\\
\frac{1}{\Vt^{3/2}} 
\log\rho
\;\;\;\;\;\;&{\rm if}\;\;\;\;{\gamma=0}
\nn\\
\,\nn \\
\frac{1}{\Vt^{3/2}} 
e^{-\rho}
\;\;\;\;\;\;&{\rm if}\;\;\;\;{\gamma=1,}
\end{array}
\right.
\ee
wherever there are strings present, and zero otherwise. 
The approximation in the transverse components 
is valid when the parallel volume changes only by a small fraction 
over the transverse directions. 

For the remaining {\bf NL}, {\bf WL}, {\bf ML} systems 
we do not know how to consider the cosmology or indeed whether 
they have any meaning at all in a cosmological setting. 
In the {\bf NL} case large scale
fluctations make it impossible to have a stable macroscopic 
system. For the {\bf WL} and {\bf ML} systems, one cannot
properly take into account the effect of winding modes. The fact that 
the canonical and microcanonical ensembles disagree in a toroidally 
compactified space
tells us that they play a dominant role in the thermodynamics, 
but it is not possible to formulate the required 
microcanonical treatment in cosmology since well defined 
global parameters such as total energy are generally lacking.

As expected $\hat{T}_0^0$ resembles the local energy density
of strings. The  $\hat{T}^i_i$ represents a relatively small 
pressure coming from Kaluza-Klein modes in the Neumann directions and 
$\hat{T}^m_m$ is a {\em negative} pressure coming from winding modes
in the Dirichlet directions. If we T-dualize the Dirichlet
directions these `winding modes'  
also become Kaluza-Klein modes in Neumann-directions and $T_m^m$ becomes 
positive. Thus negative $T_m^m$ reflects the fact that we have 
T-dualized a dimension much smaller than the string scale thereby 
reversing the pressure. For this reason negative $T_\mu^\mu$ is 
expected to be a 
general feature of space-filling excitations in transverse dimensions. 

Those who are familiar with T-duality may suspect an apparent conflict 
with this reversal of pressure when we T-dualize the extra dimensions. 
However we stress that T-duality is 
maintained. Pressure is merely {\em defined} as the change in free energy with 
{\em increase} in volume. Thus since free energy is invariant under
T-duality the pressure must reverse sign. On the other hand the 
cosmological consequences in the T-dual system must of course be the same. 
In particular gravitons do not propagate in the extra dimensions 
of the T-dualized system (\ie their Kaluza-Klein modes are heavy)
since the extra dimensions are much smaller 
than the string scale. However gravitational degrees of 
freedom come from the closed string sector in the bulk which has 
both Kaluza-Klein modes {\em and} winding modes in all directions. 
Hence in the T-dual system the light gravitational degrees of 
freedom in the extra dimensions
are winding modes. The original transverse components of Einstein's equations 
(which are valid only in the approximation that the Dirichlet
directions are 
much larger than the string scale) must in the T-dual system be replaced by 
the equations of motion derived from the action now 
expressed as a sum over gravitational {\em winding}
sectors. The upshot is that T-duality dictates that the resulting 
equations have the same consequences for the brane cosmology as the 
original system.

\section{Cosmological Equations in $D=5$}


In the previous sections we obtained the bulk stress-energy
tensor at temperatures close to the Hagedorn temperature
for the most general limiting cases in arbitrary numbers 
of dimensions. We now proceed to a discussion of the cosmology
for which we will for the most part consider a `toy-model' system with 
the following restrictions. 

The first restriction is that we implicitely be considering 
a D-brane configuration that has 3 large parallel 
dimensions (\ie the `observable
universe') and only one transverse dimension $y$ that supports winding 
modes. Such a configuration could possibly arise in a 5 dimensional 
intersection of D-branes. A more realistic case would 
probably include all $D=10$ dimensions but the Einstein equations are 
significantly more complicated in this case and we leave their 
full examination to a later paper. 
However in section 5.6 we will discuss why we expect to find the 
same behaviour in higher dimensions. 

The second restriction we make is simply for definiteness;
we assume adiabaticity 
when solving Einstein equations. We shall find
in this section and in section 5  
both power law and exponential inflation. 
{\em In principal}, as we will see in section 6, it is therefore 
possible to solve the 
horizon problem even with adiabaticity. {\em In practice} however 
it is not possible to introduce {\em a priori} the entire 
entropy of the observable universe (\ie $S\sim 10^{88}$) onto a
primordial brane of order the string scale without either 
destroying the brane or 
having an unreasonably small string coupling (\ie $g_s\sim 10^{-88}$!).
In section 6 therefore we shall discuss possible deviations from adiabaticity.

We use the metric of eq.(\ref{simple}), 
\be 
ds^2 = -n^2 dt^2 + a^2 dx^2 + b^2 dy^2 ,
\ee
which foliates the space into
flat, homogeneous, and isotropic spatial 3-planes.
Here ${\bf x} = x_1,x_2,x_3$ are the coordinates on the
spatial 3-planes while $y$ is the  coordinate
of the extra dimension.  For simplicity, we make a further
restriction by imposing $Z_2$ symmetry under $y
\rightarrow -y$.  
Without any loss of generality we choose the 3-brane
of the `observable universe' to be fixed at $y=0$.
(Even if the brane is moving we can always reparameterize the
theory so that the 3-brane is fixed in coordinate space, although
in doing so we lose further freedom of gauge choice~\cite{chuf}.)
With our assumption of homegeneity we can associate a scale 
factor with the parallel dimensions $a(t,y)$ and one for the 
`extra' transverse dimensions $b(t,y)$.
We define $a_0(t(\tau))=a(t,0)$ 
as the scale factor describing the expansion
of the 3-brane where $t(\tau) \equiv \int d\tau n(\tau,y=0)$ is the
proper time of a comoving observer.

Several authors \cite{KKOP} - \cite{yetmorerefs}  have
presented the bulk Einstein equations but for completeness we 
briefly restate the results; 
\ba
\hat{G}_{00} &=& 3\Biggl\{\frac{\dot{a}}{a}\,\Biggl(\frac{\dot{a}}{a} +
\frac{\dot{b}}{b}\Biggr) -\frac{n^2}{b^2}\,\Biggl[\frac{a''}{a} +
\frac{a'}{a}\,\Biggl(\frac{a'}{a} - \frac{b'}{b}\Biggr)\Biggr]\Biggr\}
= \hat{\kappa}^2 \, \hat{T}_{00}\,,\label{00}\\[4mm] 
\hat{G}_{ii} &=& \frac{a^2}{b^2}\Biggl\{\frac{a'}{a}\,
\Biggl(\frac{a'}{a} + 2\frac{n'}{n}\Biggr) -\frac{b'}{b}\,\Biggl(\frac{n'}{n}
+2\frac{a'}{a}\Biggr) +2 \frac{a''}{a} +\frac{n''}{n}\Biggr\}\nonumber\\[4mm]
&+& \frac{a^2}{n^2}\Biggl\{\frac{\dot{a}}{a}\,\Biggl(-\frac{\dot{a}}{a} +
2\frac{\dot{n}}{n}\Biggr) -2\frac{\ddot{a}}{a}+ \frac{\dot{b}}{b}\,
\Biggl(-2\frac{\dot{a}}{a} + \frac{\dot{n}}{n}\Biggr) -
\frac{\ddot{b}}{b}\Biggr\}= \hat{\kappa}^2\,\hat{T}_{ii}\,,
\label{ii}\\[4mm]
\hat{G}_{05} &=& 3\Biggl(\frac{n'}{n} \frac{\dot{a}}{a}
+ \frac{a'}{a} \frac{\dot{b}}{b} -\frac{\dot{a}'}{a}\Biggr)
 = \hat{\kappa}^2\,\hat{T}_{05}\,,
\label{05}\\[4mm]
\hat{G}_{55} &=& 3\Biggl\{\frac{a'}{a}\,\Biggl(\frac{a'}{a} +
\frac{n'}{n}\Biggr) -\frac{b^2}{n^2}\,\Biggl[\frac{\dot{a}}{a}\,
\Biggl(\frac{\dot{a}}{a}-\frac{\dot{n}}{n}\Biggr) +
\frac{\ddot{a}}{a}\Biggr]\Biggr\} = \hat{\kappa}^2\,\hat{T}_{55}\,, 
\label{55}
\ea
where we use the notation of ref.\cite{KKOP} in which 
\be
\hat{\kappa}^2=8\pi \hat{G}=8\pi/M_5^3,
\ee
where $M_5$ is the five-dimensional Planck mass and the dots and primes
denote differentiation with respect to $t$ and $y$, respectively.
Note that, as stated earlier, our ansatz implies that 
$T_{05}=0$ in the bulk.




The bulk equations only apply in an open region that does not include
the boundary. It is the Israel conditions which
connect the boundary and the bulk~\cite{KKOP}-\cite{yetmorerefs};
\begin{equation}
\sum_{\pm \mbox{faces}} \left( K_{\mu\nu} - K h_{\mu\nu} \right) = 
t_{\mu\nu} \, ,
\label{eq:israel}
\end{equation}
where $K_{\mu\nu}$ is the extrinsic curvature,
the sum over faces is for each side of the boundary surface and
we have defined
\begin{equation}
t_{\mu\nu}= 
\frac{2 }{\sqrt{h}} \frac{\delta S_{boundary}}
{\delta h^{\mu\nu}}
\end{equation}
as the energy momentum tensor on the boundary.  We will assume that
this energy momentum tensor on the boundary can be written in a
perfect fluid form.  For example, 
\begin{equation}
\label{eq:energydensity}
t^0_0 = \rho_{br}
\end{equation}
 and 
\begin{equation}
\label{eq:pressure}
t^1_1= - p_{br} \, ,
\end{equation}
where $\rho_{br}$ and $p_{br}$ are the energy density and
pressure, respectively, measured by a comoving observer.

We can now find the Israel conditions specific to the metric
in eq.(\ref{simple}).
Because we are considering
a brane at a $Z_2$ symmetry fixed plane, from the sum
over faces we get two identical terms,
merely resulting in a factor of two.  In a more general spacetime,
the Israel conditions would still hold, but
one would have to explicitly add the contributions from
the two sides of the boundary surfaces since these contributions
would no longer be equal.  
With the $Z_2$ symmetry, the Israel condition becomes,
\ba
3[a'/a]_0 &=& -\hat{\kappa}^2 b_0 \rho_{br} \nn\\
3[n'/n]_0 &=& \hat{\kappa}^2 b_0 (2\rho_{br}+3 p_{br}) 
\ea
The same result can be found by adding $\delta(y)t^\mu_\nu$ 
to $T^\mu_\nu$ and equating the coefficients of $\delta(y)$ in 
Einstein's equations. 

It is interesting at this point to consider the relative 
contributions of the stringy excitations and of the D-brane tension
to the cosmology. In particular we, somewhat counterintuitively, find
that the diffuse stringy component can have a dominant effect
on the cosmology even in the weak coupling limit.
To see this, first note that the effective gravitational 
coupling $\hat{\kappa}$ is given in terms of the string coupling $g_s$ by 
\be 
\frac{1}{\hat{\kappa}^2} \sim
\frac{m_s^{D-2} Vol_{D-5} }{g_s^2} \sim M_5^3 .
\ee
where $Vol_{D-5}$ is the volume of the compactification from $D=10$ to
$5$ dimensions, $m_s$ is the string scale and $M_5$ is the 
5-dimensional Planck mass. 
The intrinsic tension of a D-brane is given by 
\be 
\label{bbb}
T_{Dp}=\frac{1}{(2\pi)^p g_s} \sim \frac{1}{\hat{\kappa}\sqrt{Vol_{D-5}}}
\ee
For convenience we will henceforth assume that $ Vol_{D-5}\sim m_s^{-5}$ 
so that $g_s$ and $\hat{\kappa}$ are of the same order of
magnitude in string units (where $m_s\sim 1$). 
(This need not be the case if some of the other space dimensions
are compactified with a radius much larger or smaller than  
the string scale.)
Thus the intrinsic tension of the D-brane
satisfies $\rho_{br} \sim 1/\hat{\kappa}$,
and in the small coupling limit one might expect it to 
dominate the cosmology.  However, this is not the case.
If we substitute the Israel matching conditions 
into the cosmological equations (\ref{00}-\ref{55}), 
we see that the contributions of the 
D-brane tension and of the stringy components in the cosmological 
equations are of order  
\[
{\hat{\kappa}}^4 \rho_{br}^2 \mbox{\hspace{0.3 cm }and \hspace{0.3 cm }}
{\hat{\kappa}}^2 \frac{\rho}{\vt} \mbox{ ~,~} {\hat{\kappa}}^2\hat{p}_\gamma
\]
respectively. When the density of the string gas 
is close to the critical Hagedorn density, $\rho_c\sim 1$ (the effective 
lower bound), all the terms are of order $\hat \kappa^2 $ and are therefore 
comparable. Since the density of the string gas can be
significantly larger than the lower bound of $\rho_c \sim 1$, 
clearly the bulk stringy contribution can be significantly
larger than the brane contribution.

We can further see the dominance of the bulk stringy component
in the weak coupling limit.  All our work has assumed that
\be 
\label{bound}
\rho \leqsim T_{Dp} .
\ee
If this condition is violated, 
the thermal energy of the brane is larger than 
its rest mass and one would expect our perturbative treatment of the 
D-brane to break down (see ref.\cite{abkr} for possible outcomes.)
(Note that this bound does not apply {\it{directly} }
to orientifolds which are stuck at fixed points.)
Since $T_{Dp} \propto {1 \over \hat{\kappa}}$,
we can see that the amount of thermal energy that can be
loaded onto the brane increases as the coupling gets weaker.
When this bound is saturated, the stringy contribution to the cosmological 
equations can be ${\cal O}({\hat{\kappa}})$ while 
that from the intrinsic tension is ${\cal O}
({\hat{\kappa}}^2)$; thus weak coupling 
is advantageous for string dominance in the cosmology. This
rather surprising fact will become clear in the examples of the 
next section.

\section{Results: Behaviour of scale factors $a(t)$ and $b(t)$}

In this section we present results for the behaviour of the scale
factors $a$ and $b$. As discussed above, we will for simplicity and 
definiteness mainly consider 5 dimensional, adiabatic systems.  
In sections 5.6 and 6 we will consider how it may be possible to 
relax these restrictions.

Our results are obtained by solving 
Einstein's equations
together with the constraints provided by the Israel conditions. 
We use the energy momentum tensor derived above appropriate to a
primordial Hagedorn epoch.
First we will consider the regime {\bf L}[$\gamma$] under the assumption 
that the brane is energy/pressureless and that $\dot{b}=0$.
We also discuss some constraints imposed by the stability of the 
extra dimension. 
We then catalogue more general types of behaviour that
can occur for $\dot{b}\neq 0$ and discuss a semi-realistic
example in which $b$ is constrained.

\subsection{Hagedorn inflation; 
$\Lambda_{br} = \Lambda_{bulk} =\dot{b}=0$}

Let us first impose
$\dot b=0$ simply as an external condition; \ie
the extra dimension is fixed in time.  
We also set the cosmological constants  both on the 
brane and in the bulk to zero, 
$\Lambda_{br} = \Lambda_{bulk}= 0$. All other 
contributions to the energy momentum localized on the brane 
(for example massless Yang-Mills degrees of freedom or 
possibly an {\bf NL} subsystem) have an entropy that is subdominant 
to the limiting bulk degrees of freedom as long as 
$\rho> \rho_c$ (where $\rho_c$ is a critical density of order 1). 
For the moment we will therefore neglect them 
and set $a'(t,0)=n'(t,0)=0$.
In this discussion the brane at $y=0$ 
is playing no role in determining the evolution of the
cosmology; the scale factor $a_0$ changes purely as a result of the 
bulk equations. (We shall see shortly that consistency 
of the full solutions requires a second brane at $y=l$). 

\vspace{0.5cm}
\noindent\underline{ {\it A. Solutions to the $T_{55}$ equation}}
\vspace{0.5cm}

It is simple to solve the 55 equation
for the scale factor $a(t,y)$ at the origin, 
\be
a(t,0)=a_0(t) \, ,
\ee
for all the {\bf L}$[\gamma]$ systems.

\noindent\underline{ {\it A1.} {\bf L}[-1]{\it :}} 
In the {\bf L}[-1] systems (the high energy case with windings
in all transverse dimensions) we 
see that $T_5^5 \sim \sqrt{\rho} \sim a^{-3/2}$
(where, again, $\rho$ is the local energy density measured with 
respect to the volume $\Vp$)
and hence, 
\begin{equation}
\label{61}
a_0(t) \sim t^{4/3} \, ,
\end{equation}  
\ie {\em power law inflation}.

\noindent\underline{ {\it A2.} {\bf L}[0]{\it :}} 
We find that $\gamma=0$ gives us a
period of  
{\em exponential inflation}. To find this solution we 
write $a_0^2=\exp F$ and the $T_5^5$ equation becomes 
\[
\ddot{F} +\dot{F}^2 = \frac{2 \hat{\kappa}^2}{3} 
\left( \log{\rho(0)} - \frac{3}{2}F \right)
\]
which may easily be solved to give 
\be 
\label{expinf}
\frac{a_0(t)}{a_0(0)} = \exp \left( 
{
-\frac{t^2\hat{\kappa}^2}
{8\Vt^{\frac{3}{2}}} +t \sqrt{\frac{\hat{\kappa}^2}{6 
\Vt^{\frac{3}{2} }} \log (\rho(0)e^{3/4})}  }\right) 
\ee
where $\rho(0)$ is the initial density at $t=0$. 
Initially the second term in the exponent dominates
and there is exponential inflation.
This solution has an automatic end to inflation 
when $da_0(t)/dt \sim 0$, \ie when the two terms in the
exponent are roughly of the same magnitude. Hagedorn inflation ends at
time $t \sim\hat{\kappa}^{-1}\sqrt{\Vt^{3\over 2} \log\rho (0)}$
which corresponds to $\log \frac{a_0^3(t) }{a_0^3(0)} \approx \log \rho(0)$.
During a period of adiabaticity, the entropy and energy density
dilute as $1/a_0^3$ because the Hagedorn phase is always 
to a first approximation like pressureless matter 
($\hat{p}_{-1}\ll \hat{\rho}$). Hence we find that the above condition
for inflation to end happens 
when $\rho(t) \sim 1$ (in string units), just 
as the system is dropping out of the Hagedorn phase. 

\noindent\underline{ {\it A3. Other systems:}} 
We summarize the energy momentum tensors and cosmological 
behavior of $a_0(t)$ for the remaining systems 
in table 1, where $p=3$ in our `toy-model'. 
In all these cases it should be noted that 
during a period of adiabaticity, the entropy density 
dilutes as $1/a_0^3$. The reason for the 
various different types of behaviour is of course due to 
$T_5^5$ which may drop off more slowly than $1/a_0^3$
and in the {\bf L}[0] case drops only logarithmically with the expansion. 
Note inflationary (superluminal) expansion for $\bf{L}$[$\gamma \leq 0]$.

In adiabatic systems the Hagedorn regime and hence the inflationary 
behaviour we have found eventually come to an end.  
For a system to be in the Hagedorn regime 
requires an entropy density higher than the 
critical Hagedorn density (of order 1 in string units).
Below this density the energy momentum tensor and hence 
the cosmology is governed by the massless 
relativistic Yang-Mills gas (the gas is
present on the brane even in the Hagedorn regime but is subdominant). 
Thus there is no problem exiting 
from the inflationary behaviour.  In fact the
main issue is how long inflation can last. We return to 
this question later when we discuss how inflation can be 
sustained.

\vspace{0.5cm}
\begin{table}[ht]
{\footnotesize 
\centerline{
\begin{tabular}{|c||c|c|c|}
\hline 
regime & 
$\rho(\beta)=E/\Vp$ & 
$T_m^m = \eta' V_o/\Vt^2 \times ... $ & 
$a(t)/a(0)$  \\
\hline
\hline
&&& \\
 {\bf L}[-1] & 
$\frac{1}{\Vt}(\beta-\beta_H)^{-2}$ & 
$\rho^{\frac{1}{2}}$ & 
$t^{\frac{4}{p}}$ \\
{\bf L}[-$\frac{1}{2}$] & 
$(\beta-\beta_H)^{-\frac{3}{2}}$ & 
$\rho^{\frac{1}{3}}$ & 
$t^{\frac{6}{p}}$ \\
 {\bf L}[0] & 
$(\beta-\beta_H)^{-1}$ & 
$\log \rho$ & 
$\exp \left( {-C t^2 + D t }\right) $ \\
 {\bf L}[$\frac{1}{2}$] & 
$ (\beta-\beta_H)^{-\frac{1}{2}} $ & 
$ \rho^{-1} $ & 
$ t^{-\frac{2}{p}} $ \\
 {\bf L}[1] & 
$-\log (\beta-\beta_H)$ & 
$e^{- \rho}$ & 
const \\
&&& \\ \hline 
\end{tabular}
}}
\caption{Cosmological regimes for open strings in the Hagedorn phase
with $p$ large parallel dimensions. The constants 
$C$ and $D$ are given for a $p=3$ 
(\ie the `observable universe' as a 3-brane)
in eq.(\ref{expinf}).  Note inflationary
expansion for {\bf L}[$\gamma \leq 0$].}
\label{table1}
\end{table}

\vspace{0.5cm}
\noindent\underline{ {\it B. Solutions to the $T_{00}$ and $T_{ii}$ equations}}
\vspace{0.5cm}

We can get an approximate 
solution for the remaining equations which is valid in all the 
regions of interest as follows. For small $y$ we write 
\ba
a(t,y) &=& a_0(t) + \frac{y^2}{2} a_2(t) +\ldots\nn\\
n(t,y) &=& 1 + \frac{y^2}{2} n_2(t) +\ldots
\ea
where henceforth we normalize $n(t,0)=1$. 
First, given our constraints above, the $05$ equation 
is trivially satisfied.
As before we solve the 55 equation with
$\hat{p}_{\gamma} = \Vt^{-3/2} \rho^{2/3q}$ and find 
\be 
a_0(t) \approx At^q,
\ee
where 
\be 
b_0^{3/2} A^{2/q} = \frac{\hat{\kappa}^2 }{3 q(2q-1)}.
\ee
Solving the 00 and $ii$ equations with 
$T_0^0 = -\hat{\kappa}^2\rho/V_{\perp} = -\hat{\kappa}^2 \hat{\rho} $ 
we find 
\ba
a_2(t) &=& 
b_0^2 a_0 \left( \frac{q^2}{t^2} - \frac{\kappa^2
\hat{\rho}}{3}\right)\nn\\
n_2(t) &=& 
b_0^2 \left( \frac{q^2-2q }{t^2} + \frac{\kappa^2
(2 \hat{\rho}+3\hat{p}_{\gamma})}{3}\right).
\ea

\vspace{0.5cm}
\noindent{\underline{\it Phase diagram constraints on the 
transverse dimension and $\alpha_{||}\approx 1$}}
\vspace{0.5cm}

One important aspect of the approximate solution is that it is 
valid in all Hagedorn dominated regions. Obviously the approximation requires 
that the $y^2$ terms are small. 
Looking at $a(t,y)$ we see that the
condition is $a_2y^2 < a_0$ or 
\be 
\label{hol}
\hat{\kappa}^2 \hat{\rho} < \frac{1}{R_\perp^2}.
\ee
Recall from ref.\cite{abkr} that the Hagedorn regime 
occupies a small-coupling region between the Yang Mills phase 
and the black brane phase. Thus if
the energy density is too high, thermodynamics is dominated by 
black branes and our entire perturbative derivation ceases to be valid. 
In addition there is a `holographic' constraint 
which is saturated when the entropy is equal to that of a
black D$p$-brane filling the entire transverse volume 
(for a review see ref.\cite{oz}); 
\be 
\hat{\kappa}^2 \hat{\rho}\Vt \approx \hat{\kappa}^2 
\rho \ll \Vt^{{D-3-p}\over {D-1-p}},
\ee
where as before (see eq.(\ref{bbb})) we assume for convenience that 
the volume of the compactification to $D$ dimensions is $\sim 1$ 
so that $\hat{\kappa}\sim g_s$.
In the case that the $9-p$ transverse dimensions are 
isotropic, this is the same as eq.(\ref{hol}). The 
present case is (for convenience) taken to be one in which
there is only one transverse dimension of size $R_\perp$
which is much larger than the 
string scale. We then set $D=5$ and $p=3$ 
and again find eq.(\ref{hol}).
Thus our approximate solution for the particular case above 
breaks down precisely where the holographic bound is saturated - when 
the event horizon of a black-brane fills the transverse dimensions.

Note that, depending on the dimensionalities, the
entire calculation (\ie based on 
Hagedorn regime thermodynamics) can already break down at 
much lower densities corresponding merely to black brane dominance.
Considering for definiteness black-hole type manifolds 
with planar asympotics, $S_\beta^1\times T_L^p$, it was seen
in ref.\cite{abkr} that when 
\be
\label{eq:schwarz}
\hat{\kappa}^2\rho\gg 1 
\ee
the dominant component in the thermodynamics (\ie that phase with the 
largest entropy) is the horizon of a Schwarzschild-like black brane. 
Since $R_\perp$ is larger than the 
string scale, eq.(\ref{eq:schwarz}) can be satisfied at a density
much lower than the density of the holographic bound above.

Adding in the constraint that the Yang-Mills entropy be less than the 
Hagedorn, we find that the phase diagram restrictions can be summarized as 
\be
\mbox{min}\left\{ 
R_\perp^{D-3-p}
,1\right\}
\gg \hat{\kappa}^2\rho\gg \hat{\kappa}^2 N^2
\ee
where $N$ is the number of branes situated at $y=0$. (The additional  
requirement of $\rho\gg \rho_c\sim 1$ explains why we need to be at a small 
string coupling, $\hat{\kappa}^2 \ll 1$.)

One consequence of the phase diagram constraints is that in the 
full solution $a$ is not allowed to vary significantly
in the range $y\in [-l,l]$, and hence, as promised,
$\alpha_{||}(0)\approx 1$ in all the regions of validity. 
Note that $a''$ and $n''$ must still compensate 
for $T_0^0$ being greater than $T_5^5$ 
(the saddle-point dominance condition) and these two requirements
are compatible only at weak coupling.

\vspace{0.5cm}
\noindent{\underline{\it Stabilization and the second brane}}
\vspace{0.5cm}

A nice check of our approximate solution is that 
it should obey the stabilization
condition discussed in \cite{KKOP,E} for the case
where $T_{05} =0$ (as we are assuming throughout),
\be 
\label{stabilization}
\int dy \sqrt{g_{55}} (T_0^0 + T_i^i - 2 T_5^5 ) =0.
\ee
To see that it indeed does without any further assumptions
we need to consider what happens to the solution at 
the boundaries of the compact space.
Assume that the large transverse dimension is compactified on a 
range $[-l,l]$ and recall that for convenience 
we are assuming $Z_2$ symmetry under $y\rightarrow -y$. 
Thus at $y=l$ the solutions have a discontinuity in the 
derivatives of $a$ and $n$ which is effectively 
a {\em second} brane. Using the discontinuity condition at $y=l$,
\ba
a'_{-l}-a'_{+l} &=& -\frac{\hat{\kappa}^2}{3}\rho_{br}a(l)b_0\nn \\
n'_{-l}-n'_{+l} &=& +\frac{\hat{\kappa}^2}{3}(3p_{br} + 2 \rho_{br})
n(l)b_0,
\ea
we find that this corresponds to additional terms in the 
energy-momentum tensor given support at $y=l$ by a delta function;
\ba
\Delta T_0^0&=& 
\delta (y-l) \rho_{br}=\delta (y-l) R_\perp \left( \hat{\rho}-
3 \frac{q^2}{\hat{\kappa}^2t^2} \right) \nn\\
\Delta T_i^i&=& 
\delta (y-l) p_{br}=\delta (y-l) R_\perp \left( -\hat{p}_{\gamma}-
\frac{3 q^2 - 2 q}{\hat{\kappa}^2t^2} \right),
\ea
where $R_\perp = b_0 l$ and where 
we have neglected terms of order
$\hat{\kappa}^4 \hat{\rho}^2 {R_\perp^3}$ in accord with our 
previous discussion. Note that the signs of the brane contributions 
are the opposite of those coming from the bulk, 
so that eq.(\ref{stabilization}) now reduces to 
\be 
\label{stabilization2}
\int dy \left( \frac{2 q^2 -q}{t^2} + T_5^5 \right) =0.
\ee
This equation is automatically satisfied since it is the 
55 component of Einstein's equations. 

In the stabilized system (\ie with the $\dot{b}=0$ ansatz) the 
brane we find at $y=l$ inevitably has a 
peculiar equation of state that does not resemble a Yang-Mills
or cosmological constant on the brane. 
Nevertheless the stabilization condition 
should always be satisfied so that, in a full model 
which includes a stabilization mechanism (given our assumptions for 
$a'=b'=n'=0$ at $y=0$ and $Z_2$ symmetry), the gravitational sector 
should behave {\em as if} there were a brane with these 
characteristics located at $y=l$.

\subsection{{{\bf L}[$\gamma$]} with 
$\frac{\hat{\kappa}^2}
{12}\Lambda_{br}^2 - \Lambda_{bulk} =0$ and $\dot{b}\neq 0$}

We now study the case where the extra dimension is not stabilized 
but there is still no nett cosmological constant. 
By `nett' we mean that the contribution from $a'$ and 
$n'$ in $G_5^5$ 
cancels the bulk contribution on the RHS of the $T_5^5$ equation.
The condition for this is 
\be 
\Lambda_{nett} = -\frac{\hat{\kappa}^2}{12}\Lambda_{br}^2 + \Lambda_{bulk} =0.
\ee
In this case we could again, by studying 
the 00 and ii equations, try to find an approximate solution
which is valid throughout the Hagedorn regime as we did in the previous 
subsection. However we would 
not learn anything further by doing this since we could always independently 
adjust $a_2(t)$ and $n_2(t)$
to satisfy the equations. (There is nothing other than the phase diagram
constraints to predetermine these parameters.)  
In addition eq.(\ref{stabilization}) need not be
satisfied (and in fact the integral is $\propto \dot{b}$), but
one should keep in mind the fact that the full solution 
again has discontinuities in $a'$ and $n'$ at $y=l$ so that 
in general there should be a brane (or something that can 
compensate in a similar way) situated at $y=l$ even when $\dot{b}\neq 0$.
 The next subsection shows that by 
eliminating the second brane at $y=l$
one gains an extra constraint on $a''$ or $n''$ and 
hence useful additional information from the $ii$ and 00 equations.

Therefore we will first focus on the 55 equation.
Since $\alpha\approx 1 $, for $\gamma=-1,-\half,\half$, the pressure is 
\begin{equation}
\label{eq:t55}
\hat{p}_\gamma 
\propto a^{{-3\gamma}\over {\gamma-1}} b^{-3\over2} \, .
\end{equation}
We will now catalogue some possible types of behaviour for $a$ with 
different ans\"atze for $b$.

Taking $\Lambda_{nett}=0$, we find a family of power law 
solutions for $\gamma=-1,-\half,\half$ of the form 
\ba
\label{eq:plaw}
a_0(t) &\approx & At^q\nn\\
b_0(t) &\approx & Bt^r 
\ea 
where subscript-0 indicates values at $y=0$, 
\ba
q&=&{ {{\gamma-1}\over {2\gamma}} ({4\over 3}-r)},\nn\\
A^{{3 \gamma} \over {\gamma-1}}B^{3\over 2} &=& 
\frac{\hat{\kappa}^2}{3 q (2q-1)},
\ea
and where $r$ is arbitrary.
Note that if we choose $r=4/3$ then $b \sim t^{4/3}$
and  $a$ is stationary whatever the 
value of $\gamma$ (the complementary situation 
to $b$ stationary and $a$ undergoing power law inflation
in eq.(\ref{61})).

We also find a family of hyperbolic solutions of the form
\ba
\label{hypo}
{a_0(t)}\over{a_0(0)} &= & \left( \frac{\sinh 2C(t+t_1)}
{\sinh 2Ct_1}\right)^{1\over 2}\nn\\
{b_0(t)}\over{b_0(0)} &= & 
\left\{
\begin{array}{ll}
\left( \frac{\sinh 2C(t+t_1)}
{\sinh 2Ct_1} \right)^{\gamma\over {1-\gamma}}
\;\;\;\;\;\;&{\rm if}\;\;\;\;{\gamma=-1,-\half,\half}
  \nn\\
\;\nn\\
\log \left(
\frac{\sinh  
2C(t+t_1)}{\sinh 2Ct_1}\right)
\;\;\;\;\;\;&{\rm if}\;\;\;\;{\gamma=0}
\nn\\
\,\nn \\
\exp 
\left(
\frac{\sinh  2C(t+t_1)}{\sinh 2Ct_1}\right)
\;\;\;\;\;\;&{\rm if}\;\;\;\;{\gamma=1.}
\end{array}
\right.
\ea
where $C$ and $t_1$ are defined by 
\ba 
C^2 &=& 
 \frac{\hat{\kappa}^2}{6}\hat{p}_{\gamma}(0) 
\nn\\
{{\dot{a}_0(0)}\over{a_0(0)}} &= &  
C \coth 2Ct_1,
\ea
$\hat{p}_\gamma$ is given by eq.(\ref{eq:phat}), 
and where \eg $\hat{p}_\gamma (0),\,\rho(0) \equiv \hat{p}_\gamma
(t=0),\,\rho(t=0)$.
Note that the solutions with $t_1\rightarrow \pm\infty$ describe the
scale factor exponentially increasing or decreasing; 
\eg for $\gamma=-1,-\half,\half$ in this limit we have 
\be
\label{expo}
\begin{array}{llll}
{a_0(t)}\over{a_0(0)} &\approx & \exp \left( \pm C t\right) &
\mbox{\hspace{0.5cm},\hspace{0.3cm}}t_1\rightarrow \pm\infty \nn\\
{b_0(t)}\over{b_0(0)} &\approx & \exp \left( \pm{{2\gamma}\over {1-\gamma}} 
C t\right) &
\mbox{\hspace{0.5cm},\hspace{0.3cm}}t_1\rightarrow \pm\infty .
\end{array}
\ee
In the hyperbolic solutions $T_5^5$ is constant in time
(see eq. (\ref{eq:t55}).
Both expansion and collapse of $a$ are possible,
since $b$ compensates appropriately (in such a way as to 
keep $T_5^5$ constant). It seems rather 
paradoxical that the 
compensation can go either way depending on the sign of 
$\gamma$. For example, 
in the $\gamma=1/2$ solution of eq.(\ref{expo}), 
if $a(t)$ is expanding then, in order to 
keep $T_5^5$ constant, $b$ has to {\em expand} exponentially as well. 
The solution in the previous subsection 
for the $\dot{b}=0$, $\gamma=0$ system (section 5.1.A2, 
eq.(\ref{expinf})) is similar, although 
there is a slight difference because there $b$ is constant,
resulting in the $e^{-t^2}$ `exit' term. We should also remember
(bearing in mind our previous discussion) that {\em if adiabaticity is 
assumed} then the density $T_0^0$ 
is always diluted exponentially fast as $T_0^0\propto \exp 
\left( - C t \frac{3-\gamma}{1-\gamma}\right)$. Finally we should 
add that which, if any, of these solutions is appropriate depends on the 
initial value we choose for $\dot{b}_0$.

\subsection{{{\bf L}[$\gamma$]} with 
$\frac{\hat{\kappa}^2}
{12}\Lambda_{br}^2 - \Lambda_{bulk} \neq 0$ and $\dot{b}\neq 0$}

Next we consider the case of additional cosmological constants
in the brane and bulk. We transfer the $a'$ and $n'$ terms 
to the RHS of the 55 equation and write
\be
\label{eq:T55nett}
T_{5\, eff}^5 = - \left(\hat{p}_{\gamma}+\Lambda_{nett}\right),
\ee
where 
\be 
\Lambda_{nett}=\Lambda_{bulk}-\frac{\hat{\kappa}^2}{12}\Lambda^2_{br}.
\ee

First we can see (rather trivially) that the hyperbolic 
solutions of the previous subsection  in eq.(\ref{hypo}) 
still exist if we modify $C$ to $C'$, where
\be 
\label{newa}
C^{\prime 2} = 
\frac{\hat{\kappa}^2}{6}\left( \Lambda_{nett}+\hat{p}_\gamma(0)\right) ,
\ee
and provided $C^{\prime 2}>0$. 
Not surprisingly we recover the usual cosmological constant driven
inflation when we set $\rho=0$.
When $C^{\prime 2}<0$ (which is the case for $\Lambda_{nett} < 0$
and $|\Lambda_{nett}| > \hat{p}_\gamma(0)$), we find a singular 
collapsing solution for $a$;
\ba
\label{trigo}
{a_0(t)}\over{a_0(0)} &= & \left( \frac{\sin 2|C'|(t+t_1)}
{\sin 2|C'|t_1}\right)^{1\over 2}\nn\\
{b_0(t)}\over{b_0(0)} &= & 
\left\{
\begin{array}{ll}
\left( \frac{\sin 2|C'|(t+t_1)}
{\sin 2|C'|t_1} \right)^{\gamma\over {1-\gamma}}
\;\;\;\;\;\;&{\rm if}\;\;\;\;{\gamma=-1,-\half,\half}
  \nn\\
\;\nn\\
\log \left(
\frac{\sin  
2|C'|(t+t_1)}{\sin 2|C'|t_1}\right)
\;\;\;\;\;\;&{\rm if}\;\;\;\;{\gamma=0}
\nn\\
\,\nn \\
\exp 
\left(
\frac{\sin  2|C'|(t+t_1)}{\sin 2|C'|t_1}\right)
\;\;\;\;\;\;&{\rm if}\;\;\;\;{\gamma=1.}
\end{array}
\right.
\ea
where $t_1$ is defined by 
\be 
{{\dot{a}_0(0)}\over{a_0(0)}} =   
|C'|\cot 2|C'|t_1.
\ee

For the power law solutions eq.(\ref{eq:plaw}),
we note that the Hagedorn 
contribution to $T_5^5$ varies as $1/t^2$ and therefore decreases 
in time compared to the constant $\Lambda_{nett}$ contributions.
Thus we can consider the
55 equation in the two limits that $T_{5\, eff}^5 $ in
eq.(\ref{eq:T55nett}) is dominated 
either by the cosmological constant term or by the Hagedorn term.
For example, if $\gamma=-1,-\half,\half$, we have
\ba
\label{90a}
\left.
\begin{array}{lll}
      a_0(t) &\approx & A(t-t_0)^q \\
      b_0(t) &\approx & B(t-t_0)^r
\end{array}
\right\}
&&
T^5_{5\, eff}\approx -\hat{p}_\gamma 
\\
\label{90b}
\left.
\begin{array}{lll}
      {a_0(t)}\over{a_0(0)} &= & \left( \frac{\sinh 2C'(t+t_1)}
      {\sinh 2C't_1}\right)^{1\over 2} \\
      {b_0(t)}\over{b_0(0)} &= & \left( \frac{\sinh 2C'(t+t_1)}
      {\sinh 2C't_1} \right)^{\gamma\over {1-\gamma}}
\end{array}
\right\}
&&
|T^5_{5\, eff}| \approx |\Lambda_{nett}| \gg |\hat{p}_\gamma |,\hspace{0.1cm}
C^{\prime 2}>0  
\\
\label{90c}
\left.
\begin{array}{lll}
      {a_0(t)}\over{a_0(0)} &= & \left( \frac{\sin 2|C'|(t+t_1)}
      {\sin 2|C'|t_1}\right)^{1\over 2} \\
      {b_0(t)}\over{b_0(0)} &= & \left( \frac{\sin 2|C'|(t+t_1)}
      {\sin 2|C'|t_1} \right)^{\gamma\over {1-\gamma}}
\end{array}
\right\}
&&
|T^5_{5\, eff}|\approx |\Lambda_{nett}| \gg |\hat{p}_\gamma |,\hspace{0.1cm}
C^{\prime 2}<0
\ea
where $t_0$ is the time at which $a$ is released from its 
small initial value. 

\subsection{Bouncing universe with softened singularity}

If $\Lambda_{nett}<0$ the above solutions admit an 
oscillating universe. For example if $r< 4/3$,
we find a universe in 
which a negative cosmological constant causes 
$a$ to follow the sin curve of eq.(\ref{90c})
which 
reaches a maximum at $ |C'|(t+t_1) = \pi/4$ before heading towards zero. 
However, unlike the case of
pure negative cosmological constant in ordinary FRW, we do not 
hit a singularity because, when the scale
factor $a$ becomes small, $\Lambda_{nett}+\hat{p}_\gamma$ in eq. 
(\ref{eq:T55nett}) becomes positive and $T_5^5$ is dominated by
$\hat{p}_\gamma$.  Instead of hitting a singularity,
$a$ reaches some minimum value and rebounds upwards on the power law
solution of eq.(\ref{90a})
(with $t_0$ now marking the time of the rebound,
$\dot{a}(t_0,0)=0$). The scale factor $a$ 
follows the power law curve until the 
Hagedorn contribution to $T_5^5$ (which varies as $1/(t-t_0)^2$) drops 
below the cosmological constant piece, at which point we 
pick up the sin curve of eq.(\ref{90c}) and have 
completed a single oscillation. The oscillation 
is around the completely static solution (with $\dot{a}=\ddot{a}=\dot{b}=0$)
where the pressure $\hat{p}_{\gamma}$ is exactly compensated by a
negative cosmological constant, 
\be
\hat{p}_{\gamma}= -\Lambda_{nett}.
\ee
The behaviour of $b$ depends on the
value of $\gamma$. For $\gamma=\frac{1}{2}$ we find that 
$b$ contracts and expands with $a$. In this case,
at small $a$ we must have a power law solution with  
$r>4/3$ so that the contracting/expanding
power law solution 
matches onto a contracting/expanding $\sin$ solution. 
On the other hand for $\gamma=-1$ or $-\frac{1}{2}$,
$b$ contracts and expands out of phase with $a$, and 
we require $0<r<\frac{4}{3}$ at small $a$.
Clearly if, at small $a$, $r\approx 0$ then from eq.(\ref{90a})  
$b$ is almost static.

We stress that this singularity smoothing behaviour may or may not 
occur depending on the initial conditions (\eg $\dot{b}$) 
and additional constraints 
on the system. For example, it does occur 
in numerical solutions with $\dot{b}=0$. 

It is encouraging to see that
one contribution of stringy physics to
the cosmology can be to soften the singularity that would otherwise
appear at $a=0$. This is a familiar aspect of string theory, but an
appealing feature of the Hagedorn phase is that 
we can find it purely perturbatively.

\subsection{Full solution with an example of a `physical' brane}

In the previous subsections we discussed the cosmological behaviour 
for various ans\"atze. However, apart from the $\dot{b}=0$ constraint 
(which can be motivated by some unknown stabilization mechanism) 
the other ans\"atze are unmotivated, and we have still to show that 
the behaviour can arise in `realistic' physical systems. The reason for this 
unwanted freedom is that the system of equations is 
underconstrained -- there are 
four parameters ($a_0,~b_0,~a_2,~n_2$) but only three independent 
equation (00,~$ii$,~$55$). Therefore, we now consider an example 
in which we impose an additional constraint coming from a particular 
choice of equation of state on the brane. 
This allows us to find a full solution,
in which the behaviour observed in the previous subsection 
arises in certain limits. 

As we saw when we discussed stabilization, there is a
discontinuity in $a'$ and $n'$ which corresponds to a brane
located at $y=l$. 
Generally, the equation of state for this brane will look rather
unrealistic (as indeed it does for the static $\dot{b}=0$
solution) and we are forced to account for it by invoking 
unknown contributions from the gravitational sector.
Our first assumption is therefore that there is {\em no} second brane
(or rather, in view of the need to conserve Ramond-Ramond charge, 
that only the brane at $y=0$ has any cosmological effect).

With this assumption, continuity at $y=l$ requires $a'_l=n'_l=0$
and we must drop the approximation that the single brane at $y=0$
(our would-be universe) is `empty' (in the sense that 
$a'_0=n'_0=0$) and consider its energy 
and momentum as well. 
The Israel conditions tell us that this energy density 
is given by 
\ba
\label{zippo}
3[a'/a]_0 &=& -\hat{\kappa}^2 b_0 \rho_{br} \nn\\
3[n'/n]_0 &=& \hat{\kappa}^2 b_0 (2\rho_{br}+3 p_{br}) .
\ea
We now make an additional assumption about the equation of state 
of the brane at $y=0$; we assume it is 
also Hagedorn-like, 
\ie as well as the bulk {\bf L}[$\gamma$] degrees of freedom
there is an additional {\em localized} Hagedorn component with 
excitations near the brane. Even though
we do not understand the thermodynamic behaviour of this subsystem, we 
can be fairly confident that its energy momentum tensor 
is like pressureless matter (as all the Hagedorn systems 
are to a first approximation -- they all have $|\rho_{br} | \gg
|p_{br}|$). Note that, to simplify this example, we will neglect the 
brane tension contribution which acts like $\Lambda_{br}$.

For $p_{br}\approx 0$
it is therefore sufficient to impose the relation
\be 
\label{lucky}
n'/n = - 2 a'/a
\ee
for all $y$. Integrated, this gives
\be 
n=\frac{a_0(t)^2}{a(t)^2}.
\ee
Together with $a'_{l}=n'_{l}=0$ 
this means that the approximate solutions for $a(t,y)$ 
and $n(t,y)$ are going to be of the form
\ba
a(t,y) &= & a_0(t) + a_1(t) \left( |y|-\frac{y^2}{2l}\right) \nn\\
n(t,y) &= & 1 -2 \frac{a_1(t)}{a_0(t)} \left( |y|-\frac{y^2}{2l}\right) .
\ea
The 05 equation may easily be solved and we find 
\be 
b\,\lambda(y) =  a' a^2,
\ee
where $\lambda(y)$ is a constant of integration. 
Note that this and eq.(\ref{zippo}) imply that 
\be 
-\frac{\hat{\kappa}^2}{3}\rho_{br} = \frac{ \lambda_0 }{a_0(t)^3}
\ee
so that the brane density is indeed diluted by the 
expansion of the scale factor $a$ like 
pressureless matter.


We now apply this to the 00, $ii$ and 55 Einstein equations
which now are more constrained by the continuity at $y=l$ and by 
our choice of equation of state. Somewhat 
fortuitously the condition in eq.(\ref{lucky}) gives
a cancellation of the $a''$ and $n''$ terms and the equations at 
$y=0$ reduce to 
\ba
\frac{\dot{a}_0}{a_0}\left(\frac{\dot{a}_0}{a_0}+
\frac{\dot{b}_0}{b_0}\right)-\frac{\lambda_0^2}{a_0^6}-
\frac{\lambda_0'}{a_0^3 b_0 }
& = &\frac{\hat{\kappa}^2}{3}\hat{\rho} \nn\\
3\frac{\lambda_0^2}{a_0^6}+\frac{\dot{a}_0}{a_0}\left(\frac{\dot{a}_0}{a_0}+
2 \frac{\dot{b}_0}{b_0}\right)
+ 2 \frac{\ddot{a}_0}{a_0}
+ \frac{\ddot{b}_0}{b_0}
& = & - \hat{\kappa}^2\hat{p}_{\gamma} \nn\\
\frac{\lambda_0^2}{a_0^6}+\left(\frac{\dot{a}_0}{a_0}\right)^2+\frac{\ddot{a}_0}{a_0}
& = & + \frac{\hat{\kappa}^2}{3}\hat{p}_{\gamma} .
\ea
Dominance of the $\hat{p}$ pressure term in the 55 equation 
is only possible if 
\be 
\label{ineq}
\lambda^2_0/a_0^6 \ll (\dot{a}/a )^2 \sim 
|\hat{\kappa^2}\hat{p}_\gamma |
\ll \hat{\kappa^2}\hat{\rho}
\ee
and so the 00 equation must be dominated by the $\lambda'$ term.
Solving gives
\be 
\frac{\lambda(y)}{\lambda_0 }-1
\approx  
\frac{\rho}{\rho_{br}} \frac{y}{l}.
\ee
Consistency of our solutions therefore requires 
that $\rho_{br}\geqsim \rho$ and hence by eq.(\ref{ineq})
\be 
\hat{\kappa}^2 \rho \ll \frac{\hat{p}_\gamma}{\rho} \ll \frac{1}{\Vt}.
\ee
As for the previous solutions, this corresponds to the 
holographic constraint in the phase diagram, and implies small
coupling. Thus we again observe that,
because only ${\hat{\kappa}}^4\rho_{br}^2$ appears in the Einstein equations, 
the cosmology can be dominated by the `bulk' energy-density 
{\em even when $\rho<\rho_{br}$}, provided that the coupling is small.

For the {\bf L}[-1] systems the $ii$ and $55$ equations may be 
solved by substituting $d=\frac{a_0(t)b_0(t)}{a_0(0)b_0(0)}$ 
and $c=\frac{a_0(t)^2}{a_0(0)^2}$ so that 
\be 
\frac{\ddot{c} }{c}= \frac{2\hat{\kappa}^2}{3}\hat{p}_{-1}
\hspace{0.3cm};\hspace{0.3cm}
\frac{\ddot{d} }{d}= -\frac{4\hat{\kappa}^2}{3}\hat{p}_{-1} 
= -\frac{4\hat{\kappa}^2}{3}\hat{p}_{-1}(0)d^{-3/2}.
\ee
Defining 
\[
w=\dot{d}_0(0)^2 - \frac{16\hat{\kappa}^2\hat{p}_{-1}(0)}{3} \sqrt{d_0(0)}
\]
we find power law and hyperbolic behaviour emerging in different limits.
When 
\be 
t-t_0 \gg \frac{w^{3\over 2}}{ \hat{\kappa}^4\hat{p}_{-1}(0)^2}
\ee
we have power law behaviour (eqn.\ref{eq:plaw}) 
with 
\be 
\left(\frac{4}{3}-r\right)\left(\frac{5}{3}-2r\right) = 
\frac{1}{\hat{\kappa}^2 \hat{p}_{-1}(0)}.
\ee
So for large $\hat{p}_{-1}$, $r\approx 4/3$, $5/6$. 
On the other hand when 
\be 
\label{cond}
t-t_0 \ll \frac{w^{3\over 2}}{ \hat{\kappa}^4\hat{p}_{-1}(0)^2}
\ee
we have $\sqrt{d}\approx - w /{ \hat{\kappa}^4\hat{p}_{-1}(0)^2} 
\approx \sqrt{d(0)}$ and hence we find the hyperbolic solution 
of eq.(\ref{90b}) with expanding $a$ and collapsing $b$. 

\subsection{More dimensions}

Finally we briefly remark on the extension to higher numbers 
of dimensions. In type I,IIA/B models the total number of 
dimensions is $D=10$ and the energy momentum tensor
is therefore of the form $T_\mu^\nu = 
diag(T_0^0,\hat{p}_{\gamma},\hat{p}_{\gamma}\ldots, 
-\hat{p}_{\gamma},-\hat{p}_{\gamma}\ldots, 0\ldots,0)$. A D$p$-brane has
$9-p$ large Dirichlet directions giving 
\be
-1 \leq \gamma \leq \mbox{min}\left\{1,\frac{7-p}{2}\right\} \, .
\ee
As we discussed in the introduction, it is always true that
$\gamma \leq 1$ because
for $p< 5$, the only possible limiting system has
excited winding modes and is the $\gamma=-1$ system.

A diagonal metric with universal scale factors is 
consistent with the above form of $T_\mu^\nu$. We choose the metric 
to be a function of a single transverse dimension which we label $y$, 
and add $D-2-p$ additional transverse dimensions 
labelled $y_m$, ($m=p+2,..D$); 
\be 
d s^2 = -n^2 dt^2 + a(t,y)^2 dx_i^2 + b(t,y)^2 dy^2 + d(t,y)^2 dy_mdy_m .
\ee
This metric gives 4 independent equations (00,$ii$,$55$,$mm$) for the 
higher dimensional systems.  Most importantly, $a''$, $n''$ and $d''$ 
do not appear in the 55 components of $G_\mu^\nu$~\cite{cedric}.
Consequently the 55 equation may again be treated separately from the 
$00$, $ii$, $mm$ equations.  The latter 3 equations 
may always be solved independently
(as in the $D=5$ case studied in detail in this paper) 
using the three unknowns $a''$, $n''$ and $d''$. 
Thus, although we will not discuss the higher dimensional case
in detail here, we can immediately see on dimensional grounds that the 
solutions to the $55$ equations with $\dot{d}=\dot{b}=0$ must be 
as shown in table 1 (with the appropriate value of $p$). 
An amusing aspect of this is that power law
inflation in {\bf L}[-1] systems (\ie the universal high energy system)
requires $p\leq 3$ regardless of the total number of dimensions. 

\section{Sustaining inflation and solving the Horizon Problem}

One of the most striking features of the cosmological solutions we have 
found is that they automatically predict a superluminal
growth of the scale factor
which is usually an ingredient of
standard inflation. The remaining point that needs to be addressed is that
our analysis assumed adiabaticity, and in order to realistically 
solve any cosmological problems, this assumption (as we will see)
almost certainly requires modification.
In this section we discuss solving the horizon problem
beginning with the causality
condition and then addressing the issue of adiabaticity.
We then discuss how nonabiabatic effects may contribute to a sustained
period of inflation.

\subsection{Causality Condition:}

An estimate of the size of the observable universe today
is given by the distance light could travel between photon
decoupling and now, 
\begin{equation}
\label{eq:dobs}
d_{obs} \sim a(t_o) \int_{t_{dec}}^{t_o}
dt/a(t) \, .
\end{equation}
Note that $d_{obs} = {\cal O}(1) \times (t_o - t_{dec})$ for $a \propto t^p$
and $p={\cal O}(1)$ between $t_{dec}$ and $t_o$.  Here $a$ is the
scale factor. We can compare the comoving size of the observable
universe to the comoving size of a causally connected region at
some earlier time $t_p$: 
\begin{equation}
\label{eq:dhor}
d_{hor}(t_p)/a(t_p)
= \int_0^{t_p} {dt\over a(t)} \, .
\end{equation}  
The observable universe today
fits inside a causally connected region at $t_p$ if
\begin{equation}
{d_{hor}(t_p) \over a(t_p)} \geq {d_{obs} \over a(t_o)} \, .
\label{eq:causality}
\end{equation}
Here, subscript-$o$ refers to today and
subscript-$p$ indicates some primordial time before the 
inflationary period of interest. 
If condition \eqr{eq:causality} is met, then the horizon size
at $t_p$ (before nucleosynthesis) is large enough to allow for a
causal explanation of the smoothness of the universe today.  Note that
more creative explanations of large scale smoothness may not involve
comparing these two patches.
For example, in the context of the brane scenarios, one might imagine
that two regions of our observable universe which seem to be causally
disconnected might in fact have talked to each other because of a geodesic
between them that went off our brane, into the bulk, and then back
onto our brane at some distant point, as demonstrated
by Chung and Freese \cite{chuf2}. In the remainder of our discussion
here we restrict
ourselves to the case where \eqr{eq:causality} is relevant; this is
certainly the case for all the brane and boundary inflation models
considered to date.

For power law expansion of the scale factor both before $t_p$ and
after $t_{dec}$ (which may or may not be the case), we can take
$t \sim H^{-1}$ during these periods. The causality condition
\eqr{eq:causality} then becomes
\begin{equation}
{1 \over a_p H_p} \geq {1 \over H_o a_o} \, ,
\label{eq:causality2}
\end{equation}
or, equivalently,
\begin{equation}
\label{eq:causality3}
{t_p \over a_p} \geq {t_{today} \over a_{today}} \, .
\end{equation}
We take the Hubble constant today to be 
\begin{equation}
H_o = \alpha_o^{1/2}T_o^2/m_{pl}(t_o),
\label{eq:htoday}
\end{equation}
where $\alpha_o = (8 \pi/3)(\pi^2/30)g_*(t_o)\eta_o$,
$g_*$ is the number of relativistic degrees of freedom and
$\eta(t_o) \sim 10^4-10^5$ is the ratio today of the energy
density in matter to that in radiation.
From eq.(\ref{eq:causality2}) we can then see that
accelerated expansion of the scale factor with $\ddot a >0$ is
required to solve the horizon problem.
As we have shown throughout the paper , such accelerated
expansion can easily occur during the Hagedorn regime.

We must here assume that, within an intitial horizon patch
of size $a_p$, there has been enough smoothing that
it is sensible to talk about a single value of the temperature
and entropy density.  This same assumption must be made
in every inflation model.

\subsection{The issue of non-adiabaticity}

Adiabatic Hagedorn inflating systems can 
{\em in principle} already solve the horizon 
problem (as opposed to standard inflation in which
nonadiabaticity is required). 
These systems could start with a large initial entropy due to
the proximity of the initial temperature to the Hagedorn temperature.
Subsequently, the entropy remains constant as the temperature drops
and the scale factor $a$ grows by a large amount.  However
{\em in practice} it is difficult to solve the horizon problem 
assuming adiabaticity: 
it is not possible to introduce {\em a priori} the entire 
entropy of the observable universe (\ie $S\sim 10^{88}$) onto a
primordial brane of order the string scale without either 
destroying the brane or 
having an unreasonably small string coupling (\ie $g_s\sim 10^{-88}$!).

Consider for example the ${\bf L}[-1]$ system with $a \sim t^{4/3}$.
Then from the first entry in Table I, we see that the entropy
on a world-volume $a_p^3$ is
\begin{equation}
\label{eq:priment}
S_p \sim \beta_H a_p^3 \rho \sim \beta_H a_p^3
{ 1 \over (\beta -\beta_H)^2} { 1 \over \vt }.
\end{equation}
We will assume that the entropy on the 3-brane today is
$S\propto (a_{today} T_{today})^3$.  

We obtain the temperature/time relation during the
Hagedorn phase from $\rho \sim a^{-3} \sim t^{-4}$ by equating
this expression for $\rho$ with that from the first entry
in Table I.  We find $t_p = \vt^{1/4} \sqrt{\beta_p - \beta_H}$.
Eq.(\ref{eq:causality}) can therefore be satisfied if
\be 
\label{eq:ratios}
\frac{S_0}{S_p} > \frac{m_s}{T_{today}} \left( 
\vt (\beta_p-\beta_H)^2 \right)^{1/4} = \frac{m_s}{T_{today}}
\rho_p^{-1/4}.
\ee
Hence even an adiabatic system here can in principle solve the horizon
problem: provided that the primordial density is large enough (or 
the primordial temperature sufficiently close to $T_H$) the 
horizon problem is solved even if $S_0=S_p$. 
Indeed, we can see from eqn.(\ref{eq:priment}) that the initial 
entropy density can be as large as we like provided 
that we are prepared to accept a $\beta $ that 
is arbitrarily close to $\beta_H$.

The serious practical 
difficulty arises however when we consider the stability 
bound in eq.(\ref{bound}) since it clearly implies that 
non-perturbative effects will be important unless the string coupling 
is fantastically weak. 
Indeed, generally if one assumes adiabaticity 
in attempting to solve the horizon problem, one
requires the entire entropy of the observable universe to be present 
in string excitations at $t=t_p$.  Hence, if the volumes are initially 
of order the string scale, then the energy density must be enormous, 
$\rho(t_p)\sim 10^{88}$ (e.g. in eqn.\ref{eq:ratios}). 
After 60 $e$-folds of inflation $\rho(t)$ falls below 
$\rho_c$, the critical density
which is needed to be in the Hagedorn phase, and the 
Yang-Mills phase takes over. 
However, for such a large
initial value of the energy density, brane stability at $t=t_p$ 
(\ie the constraint in eq.(\ref{bound})) 
requires an initial value of $g_s < 10^{-88}$.

Because of this it is worth looking more critically
at the supposed adiabaticity. This is a crucial 
assumption that almost certainly 
requires modification in a de Sitter background since the latter
possesses a horizon with its own associated entropy. 
Hence, even if adiabaticity were the correct criterion (which it 
is not), it would only be meaningful for the coupled 
string/de Sitter system since there is a backreaction of the 
horizon on the string gas. The upshot for strings in a de Sitter 
background is that they are unstable to fluctuations and 
that this instability can sustain a period of de Sitter 
inflation~\cite{englert,turok,veneziano}. This phenomenon is well known 
for strings once they are in the de Sitter-like 
phase\footnote{We qualify `de Sitter' because the exponential 
solutions we have do not possess the full O(1,4) de Sitter symmetry.}. 
However the missing ingredient that the present study adds 
is an explanation for how the universe enters a de Sitter like phase 
in the first place.   Our mechanism gets the universe into
the locally de Sitter phase, whereupon the mechanism of
refs.\cite{englert,turok,veneziano} keeps it there 
{\em without} having to assume adiabaticity.

To achieve a sustained period of inflation therefore, we could just 
invoke the findings of refs.\cite{englert,turok,veneziano}. However
for the remainder of this section we briefly elaborate on this 
property of strings, using purely heuristic arguments, 
in order to indicate a possible direction for future study.

Let us return to the density of states and consider instead 
a truly static metric in which the relevant Killing field is
the time-translation,
\be 
K^\mu \equiv \frac{\partial }{\partial t}.
\ee
De Sitter space is an example of a static metric;
in an appropriate coordinate system the metric is 
\be 
ds^2 = -dt^2 (1-H^2 r^2 ) + \frac{1}{1-H^2r^2 }dr^2 + r^2 d\Omega^2 
\ee
and the Killing field has components $(1,0,0,0)$.
We can therefore discuss thermodynamics by compactifying 
on an imaginary time coordinate with $it \equiv it+\beta$. 

A suitable
modification to discuss the present case 
is to add an additional $y$ coordinate with 
a metric $g_{55} = b(y)^2$ which is time independent 
in {\em this} coordinate system and also independent of $r$;
\be 
ds^2 = - (1-H^2 r^2 ) dt^2 + \frac{1}{1-H^2r^2 }dr^2 + r^2 d\Omega^2 +
b(y)^2 dy^2.
\ee
Under this assumption each 4 dimensional leaf of the 
foliated space is a de Sitter space and the geometry 
unambiguously fixes the Hawking temperature of the 
Horizon, $2\pi /H = \beta $.

It is well established (see for example ref.\cite{brustein} and
references therein)
that the combined (geometric plus thermal) 
system obeys a generalized second law which 
is that the total entropy,
\be 
S= S_{matter} + S_{horizon}=  S_{matter} + 
\frac{1}{4 \hat{\kappa}^2 } A_{horizon},
\ee
obeys $\delta S>0$, where $A_{horizon}$ is the area of the horizon. 
The constant-time surface $\Sigma $ extends to the horizon 
$r \approx 1/H  = \beta/2\pi $ so that 
\be 
V_{||}=Vol(S_{d_{||}}) r^{d_{||}} \sim \beta ^{d_{||}} ; A_{horizon}=
Vol(S_{d_{||}-1}) r^{d_{||}-1} \sim \beta^{d_{||}-1}
\ee
where $Vol(S_{d_{||}})$ is the volume of the unit sphere in 
$d_{||}$ dimensions. 

Thus, assuming equilibrium 
at the temperature $\beta^{-1}$, the total entropy of 
an {\bf L}[-1] string gas in a de Sitter background is of 
the form
\be 
S\sim \frac{\beta^{d_{||}}}{\vt (\beta-\beta_H)^2}  
+ \frac{\beta^{d_{||}-1}}{4\pi 
\hat{\kappa}^2}.
\ee
This function has a minimum at
\be 
T_{crit}= T_H \left(1-{\cal O}\left( 
(\hat{\kappa}^2 /\vt )^{1/3}\right)\right).
\ee
Below $T_{crit}$ fluctuations tend to drive the universe 
towards low temperature, \ie out of the de Sitter 
phase\footnote{We recognize that, in making this heuristic argument, 
we are on rather thin ice since we are simultaneously 
assuming a string coupling which is small enough for our approximations
to be correct, but large enough to maintain equilibrium between the 
different degrees of freedom.}. 
The horizon has a negative specific heat which becomes relatively 
larger as the temperature drops so that equilibrium can 
be maintained (the condition being $|C_{V-}| > C_{V+}$).
Above $T_{crit}$ in the presence of a thermal gas of strings
the flow is in the opposite direction towards an asymptotically limiting 
Hubble constant, $H_{asymp} = 2\pi T_H $. 
Thus the generalized second law indicates that fluctuations 
in the geometry tend to drive the universe even further into 
the de Sitter phase provided that the Hubble constant is 
greater than the critical value.
The entropy that the horizon loses when the Hubble constant is 
increased is more than offset by the huge increase in string entropy. 
Moreover, since the specific heat of the string gas becomes larger 
when the temperature increases, inevitably
equilibrium is lost and energy flow to the strings becomes 
apparently limitless.

There is a slight difference between this picture and 
the results of ref.\cite{turok} which we should comment on. 
In ref.\cite{turok}, the density never goes {\em above}
a critical density, $\rho'_c$,
and indeed asymptotes to it from below as one goes back in time. 
In our case, on the other hand, we must have $\rho > \rho_c \sim 1$ for the 
calculation to be valid (\ie to be in the Hagedorn regime).
Indeed in the present paper the energy density diverges 
as we approach the Hagedorn temperature. The difference 
is because ref.\cite{turok} was concerned with 
{\em stretched} strings and 
introduced a cut-off in the momentum integral
in order to find the fractal dimension of the string. 
This `coarse graining' put an artificial
upper limit on the amount of string that can be packed into the 
volume. In fact they found the maximum fractal dimension (\ie 2) only
at the critical density $\rho'_c$. A fractal dimension of 2 is typical of the
random walk behaviour which exists once Hagedorn behaviour sets in. 
So the coarse graining in ref.\cite{turok} 
effectively removed the Hagedorn behaviour and consequently 
ref.\cite{turok} found that
$\rho=\rho'_c$ gave $\beta=\beta_H $.
Conversely, the present paper begins in the regime $\rho > \rho_c$ 
where the calculations of ref.\cite{turok} end.


\section{Conclusion and discussion}

In this paper we have studied the possible cosmological implications of
the Hagedorn regime of open strings on D-branes in the weak coupling limit. 
Our main result in sections 2-5 is that, due to the non-extensive
dependence of the free energy on the volumes, 
a gas of open strings can exhibit negative pressure 
leading naturally to a period of power law or even exponential
inflation -- Hagedorn inflation.
We also find that the open string gas can dominate the cosmological 
evolution at weak coupling even though the D-brane tension becomes
large in this limit.

Hagedorn inflation also has a natural exit 
since any significant cooling can cause a change in the thermodynamics
if winding modes become quenched or if the density drops 
below the critical density, $\rho_c\sim 1$, needed for the entropy 
of the Hagedorn phase 
to be dominant. Such a cooling can be caused by a sudden adiabatic
increase in the transverse radius or by the inflation itself. 
We find this `easy-exit' feature of 
open-string Hagedorn inflation to be of its most appealing features. 
In addition, we found that a small but negative cosmological constant,
can cause the universe to enter 
a stable but oscillating phase.  The effect of
the Hagedorn phase is to soften the singular behaviour associated
with the bounce. 

The most striking aspect of our discussion is probably the 
existence of negative pressure. One might therefore ask 
how general a feature this is expected to be. By T-dualizing
we argued that we can put the negative pressure down to the fact that 
in any particular direction, the gravitational degrees of 
freedom have both Kaluza-Klein modes 
and winding modes whereas the open strings (which are dominant in the 
entropy) have only one or the other.  
Hence we expect negative pressure to be possible 
whenever there are large space-filling modes that 
dominate the entropy.

In section 6 we speculated on how the inflation might be 
sustained through the well known phenomenon of 
string instability in a de Sitter background. 

Hagedorn inflation may be thought of as a first example in the 
search for alternatives to the cosmological constant 
within the framework of string/brane systems.
One possible direction for further study in this 
area is connected with the fact that we have throughout 
been taking the string coupling to be weak enough so 
that the brane tension does not play an important role in the 
cosmology. It is therefore interesting to ask if new 
cosmological effects might arise from large scale fluctuations in the
branes themselves~\cite{abkr,riotto}.
The arguments of section 6 indicate that if it
does then the brane driven inflation
may be qualitatively different to the string driven inflation discussed here.
This is because open strings are one dimensional objects 
with $S\sim E + const \sqrt{E} $ whereas  
fluctuating $p$-branes have an entropy~\cite{penalba}
\be 
S\sim E^{{2p} \over{p+1}}.
\ee
On calculating $\beta=\partial S/\partial E$ we see that 
$p\neq 1$ branes do 
not have the divergent behaviour which is the defining 
feature of strings. The 
thermodynamics of these objects has been the subject of much study
\cite{penalba} and there is probably more to be learnt here. 

Finally, an interesting connected issue which we did not discuss is
related to the effect of brane melting discussed in ref.\cite{abkr}, in
which the non-perturbative aspect of D-brane thermal production must be
taken into account.  At the present time it is hard to  make any 
quantitative estimate of these effects on the cosmological solutions 
we have been discussing here, but we hope to be able to address
this question in future work.

\subsection*{Acknowledgements}

\noindent We thank Dan Chung, Cedric Deffayet, Emilian
Dudas, Keith Olive, Geraldine Servant, Carlos Savoy and 
Richard Woodard for discussions. 
S.A.A. and I.I.K. thank 
Jose Barb\'on and Eliezer Rabinovici for a previous collaboration
and discussions concerning this work.
S.A.A. thanks the C.E.A. Saclay for support.
K.F. acknowledges support from the Department of
Energy through a grant to the University of Michigan.  K.F. thanks
CERN in Geneva, Switzerland and the Max Planck Institut fuer Physik in
Munich, Germany for hospitality during her stay. 
I.I.K. is  supported in part by PPARC rolling grant
PPA/G/O/1998/00567, the EC TMR grant FMRX-CT-96-0090 and  by the INTAS
grant RFBR - 950567. 



\begin{thebibliography}{99}
\frenchspacing
\def\prpts#1#2#3{Phys. Reports {\bf #1}, #2 (#3)}
\def\prl#1#2#3{Phys. Rev. Lett. {\bf #1}, #2 (#3)}
\def\prd#1#2#3{Phys. Rev. D {\bf #1}, #2 (#3)}
\def\prc#1#2#3{Phys. Rev. C {\bf #1}, #2 (#3)}
\def\plb#1#2#3{Phys. Lett. {\bf #1B}, #2 (#3)}
\def\npb#1#2#3{Nucl. Phys. {\bf B#1}, #2 (#3)}
\def\apj#1#2#3{Astrophys. J. {\bf #1}, #2 (#3)}
\def\apjl#1#2#3{Astrophys. J. Lett. {\bf #1}, #2 (#3)}

\bibitem{guth}
A. Guth, \prd{23}{347}{1981}

\bibitem{carlitz}
R. Hagedorn, {\it Suppl.~Nuovo Cimento} {\bf 3} (1965) 147;
S. Frautschi, \prd{3}{2821}{1971}; 
R.D. Carlitz, \prd{5}{3231}{1972}

\bibitem{general}
K. Huang and S. Weinberg,
\prl{25}{895}{1970};
E. Alvarez, \prd{31}{418}{1985}; \npb{269}{596}{1986}; 
M. Bowick and L.C.R. Wijewardhana, \prl{54}{2485}{1985};
B. Sundborg, \npb{254}{883}{1985};
S.N. Tye, \plb{158}{388}{1985};
P. Salomonson and B. Skagerstam, \npb{268}{349}{1986}; {\it Physica}
{\bf A158} (1989) 499; 
E. Alvarez and M.A.R. Osorio, \prd{36}{1175}{1987}; 
D. Mitchell and N. Turok, \prl{58}{1577}{1987}; \npb{294}{1138}{1987};  
I. Antoniadis, J. Ellis and D.V. Nanopoulos, \plb{199}{402}{1987}; 
M. Axenides, S.D. Ellis and C. Kounnas, \prd{37}{2964;}{1988}; 
I.I. Kogan, JETP. Lett. {\bf 45} (1987) 709;
B.Sathiapalan,  \prd{35}{3277}{1987};  
J. Atick and E. Witten, \npb{310}{291}{1988};
A.A. Abrikosov Jr. and Ya. I. Kogan
\ijmpa{6}{1501}{1991} (submitted 1989),
{\it Sov. Phys. JETP} {\bf 69} (1989) 235; 
R. Brandenberger and C. Vafa, \npb{316}{391}{1989};
M.J. Bowick and S.B. Giddings, \npb{325}{631}{1989};
S.B. Giddings, \plb{226}{55.}{1989};
R. Brandenberger and C. Vafa, \npb{316}{391.}{1989};
F. Englert and J. Orloff, \npb{334}{472}{1990}; 
B.A. Campbell, J. Ellis, S. Kalara, D.V. Nanopoulos, K.A. Olive, 
\plb{255}{420}{1991};
B.A. Campbell, N. Kaloper, K.A. Olive, 
\plb{277}{265}{1992};
S.A. Abel, \npb{372}{189}{1992};
N. Kaloper, K.A. Olive, {\it Astropart.Phys.}{\bf 1} (1995) 185;
N. Kaloper, R. Madden, K.A. Olive, \plb{371}{34,}{1996},
\hepth{9510117};
N. Kaloper, R. Madden, K.A. Olive, \npb{452}{677,}{1995} 
\hepth{9506027};
M.L. Meana, M.A.R. Osorio and  J.P. Penalba,  
\plb{400}{275,}{1997}  
\hepth{9701122}; \plb{408}{183}{1997}   
\hepth{9705185};     
K.R. Dienes, E. Dudas, T. Gherghetta and A. Riotto, \npb{543}{387}{1999}

\bibitem{deo}
N. Deo, S. Jain and C.-I. Tan, \plb{220}{125}{1989};
\prd{40}{2646}{1989}; N. Deo, S. Jain, O. Narayan and C.-I. Tan,
\prd{45}{3641}{1992}

\bibitem{thorl}
D.A. Lowe and L. Thorlacius, \prd{51}{665}{1995}, \hepth{9408134};
S. Lee and L. Thorlacius, \plb{413}{303}{1997}, \hepth{9707167}

\bibitem{abkr}
J.L.F. Barb\'on, I.I. Kogan and E. Rabinovici, \npb{544}{1999}{104}, 
\hepth{9809033}; S.A. Abel, J.L.F. Barb\'on, I.I. Kogan, and E. Rabinovici,
JHEP {\bf 04} (1999) 015

\bibitem{after}
M.L. Meana and J.P. Penalba,
\npb{560}{154}{1999}; \plb{447}{59}{1999};
M.A. Vazquez-Mozo, \prd{60}{106010}{1999};
B. Sundborg, \hepth{9908001};
S.A. Abel, J.L.F. Barb\'on, I.I. Kogan, and E. Rabinovici,
\hepth{9911004}

\bibitem{polch}
For reviews see: J. Polchinski, {\em TASI} lecturs on D-branes, 
\hepth{9611050}; ``String Theory'', vols 1,2 (CUP) 1998; C. Bachas,
``Lectures on D-branes'', \hepth{9806199}

\bibitem{inprep}
S.A. Abel and E. Dudas, in preparation

\bibitem{KKOP} P. Kanti, I. I. Kogan, K. A. Olive, M. Pospelov,
Phys.Lett. B468 (1999) 31, \hepph{9909481}; \hepph{9912266}

\bibitem{moreref} 
K. Benakli, Int. J. Mod. Phys. {\bf D8} (1999) 153,
\hepth{9804096};
N. Arkani-Hamed, S. Dimopoulos and G. Dvali,
Phys. Rev. {\bf D59}, 086004 (1999), \hepph{9807344};
H.S. Reall,  \prd{59}{103506,}{1999}
\hepth{9809195};
N. Kaloper and A. Linde, Phys. Rev. {\bf D59} (1999) 101303, 
\hepph{9811141};
G. Dvali and S.H.H. Tye,
\plb{450}{72}{1999}, \hepph{9812483};
A. Lukas, B.A. Ovrut and D. Waldram, \hepth{9902071};
T. Banks, M. Dine and A. Nelson,
\hepth{9903019};
H.A. Chamblin and H.S. Reall, \npb{562}{133}{1999}, \hepth{9903225};
N. Arkani-Hamed, S. Dimopoulos, N. Kaloper and J. March-Russell,
\hepph{9903224}; \hepph{9903239};
P. Bin\'etruy, C. Deffayet and D. Langlois, \hepth{9905012};
N. Kaloper, Phys. Rev. {\bf D60} (1999) 123506, \hepph{9905210};
T. Nihei, Phys. Lett. {\bf B465} (1999) 81, \hepph{9905487};
C. Cs\'aki, M. Graesser, C. Kolda and J. Terning, Phys. Lett.
{\bf B462} (1999) 34, \hepph{9906513}; 
J.M. Cline, C. Grojean and G. Servant, Phys. Rev. Lett.
{\bf 83} (1999) 4245, \hepph{9906523}

\bibitem{chuf}
D. Chung and K.T. Freese, \prd{61}{023511}{2000}, \hepph{9906542}

\bibitem{gold}
W.D. Goldberger and M.B. Wise, Phys. Rev. {\bf D60} (1999)
107505, \hepph{9907218}; 
\prl{83}{4922}{1999}, \hepph{9907447};
H.B. Kim and H.D. Kim, \prd{61}{064003}{2000}, \hepth{9909053}

\bibitem{E} U. Ellwanger, \hepth{9909103}

\bibitem{yetmorerefs}
T. Shiromizu, K. Maeda and M. Sasaki, \grqc{9910076};
C. Grojean, J. Cline and G. Servant, \hepth{9910081};
P. Kraus, \hepth{9910149};
P. Bin\'etruy, C. Deffayet, U. Ellwanger and D. Langlois,
\hepth{9910219};
A. Kehagias and  E. Kiritsis, JHEP 9911 (1999) 022;
E. Flanagan, S.H.H. Tye and I. Wasserman, \hepph{9910498};
D. Ida, \grqc{9912002};
 C. Cs\'aki, M. Graesser, L. Randall, and J. Terning,
\hepph{9911406};
W.D. Goldberger and M.B. Wise, \hepph{9911457}

\bibitem{oz}O.~Aharony, S.~S.~Gubser, J.~Maldacena, H.~Ooguri and Y.~Oz,
\hepth{9905111}

\bibitem{cedric}
C. Deffayet, private communication

\bibitem{chuf2}
D. Chung and K.T. Freese,
``Can Geodesics in Extra Dimensions Solve the Horizon
Problem?" \hepph{9910232}

\bibitem{englert}
R. Brout, F. Englert and E. Gunzig, Ann. Phys. {\bf 115} (1978) 78;
Gen. Rel. Grav. {\bf 10} (1979) 1; R. Brout, F. Englert and P. Spindel, 
\prl{43}{417}{1979}; Y. Aharanov and A. Casher, \plb{166}{289}{1986};
Y. Aharonov, F. Englert and J. Orloff, \plb{199}{366}{1987}

\bibitem{turok}
N.Turok, \prl{60}{549}{1988}

\bibitem{veneziano}
N.~Sanchez and G.~Veneziano,
Nucl.\ Phys.\  {\bf B333} (1990) 253;
M.~Gasperini, N.~Sanchez and G.~Veneziano,
Int.\ J.\ Mod.\ Phys.\  {\bf A6} (1991) 3853;
Nucl.\ Phys.\  {\bf B364} (1991) 365;

\bibitem{brustein}
R.~Brustein, S.~Foffa and R.~Sturani,
Phys.\ Lett.\  {\bf B471} (2000) 352,
\hepth{9907032};
R.~Brustein,
\grqc{9904061}

\bibitem{riotto}
A. Riotto,
\hepph{9904485}

\bibitem{penalba}
E. Alvarez, T. Ortin, Mod.Phys.Lett.{\bf A7} (1992) 2889;
A.~A.~Bytsenko, K.~Kirsten and S.~Zerbini,
Phys.\ Lett.\  {\bf B304} (1993) 235;
A.A. Bytsenko and S.D. Odintsov,
Prog.\ Theor.\ Phys.\  {\bf 98} (1997) 987;
I.R. Klebanov and A.A. Tseytlin,
Nucl.\ Phys.\  {\bf B479} (1996) 319;
J.P. Penalba,
\npb{556}{152}{1999}

\end{thebibliography}
\end{document}